\documentclass[aps,prl,showkeys,twocolumn,superscriptaddress,
	      floats,floatfix,preprintnumbers]{revtex4}
\usepackage{lipsum}
\usepackage{soul}
\usepackage{amssymb,amsmath}
\usepackage{hyperref}
\usepackage{xr}
\usepackage{ulem}
\usepackage{siunitx}
\usepackage{bm}
\usepackage{graphics}
\usepackage{subfigure}
\usepackage{epsfig}
\usepackage{ulem}
\usepackage{color}
\usepackage{soul}
\usepackage{algpseudocode}
\sethlcolor{yellow}
\setulcolor{blue}
\setstcolor{red}
\definecolor{darkgreen}{rgb}{0,0.625,0}
\usepackage{booktabs}
\usepackage{float}
\usepackage{placeins}
\PassOptionsToPackage{hyphens}{url}\usepackage{hyperref}

\makeatletter
\newcommand*{\addFileDependency}[1]{
  \typeout{(#1)}
  \@addtofilelist{#1}
  \IfFileExists{#1}{}{\typeout{No file #1.}}
}

\newcommand*{\myexternaldocument}[1]{%
    \externaldocument{#1}%
    \addFileDependency{#1.tex}%
    \addFileDependency{#1.aux}%
}

\myexternaldocument{RSI}

\renewcommand{\theequation}{\arabic{equation}}

\begin{document}

\title{Emergent Strategies for Shepherding a Flock}

\author{
Aditya Ranganathan$^{1}$, Dabao Guo$^{1}$, Alexander Heyde$^{2}$, Anupam Gupta$^{4}$, L. Mahadevan$^{1,2,3}$}

\address{$^{1}$School of Engineering and Applied Sciences, Harvard University, Cambridge, MA 02138 \\
$^{2}$Department of Organismic \& Evolutionary Biology, Harvard University, Cambridge, MA 02138\\
$^{3}$Department of Physics, Harvard University, Cambridge, MA 02138\\
$^{4}$Department of Physics, Indian Institute of Technology, Hyderabad 502284}



\begin{abstract}
We investigate how a shepherd should move to effectively herd a flock towards a target. Using an agent-based (ABM) and a coarse-grained (ODE) model for the flock, we pose and solve for the optimal strategy of a shepherd that must keep the flock cohesive and  coerce it towards a target. Three distinct strategies emerge naturally as a function of the scaled herd size {and} the scaled  shepherd speed: (i) mustering, where the shepherd circles the herd to ensure compactness, (ii) droving, where the shepherd chases the herd in a desired direction while sweeping back and forth, and (iii) driving, where the flock surrounds a shepherd that drives it from within. A minimal dynamical model for the size, shape, and position of the herd captures the effective behavior of the ABM and further allows us to characterize the different herding strategies in terms of the behavior of the shepherd that librates (mustering), oscillates (droving), or moves steadily (driving).
\end{abstract}

\maketitle

\section{Introduction}

\begin{figure*}
\centering
\includegraphics[width=1.0\linewidth]{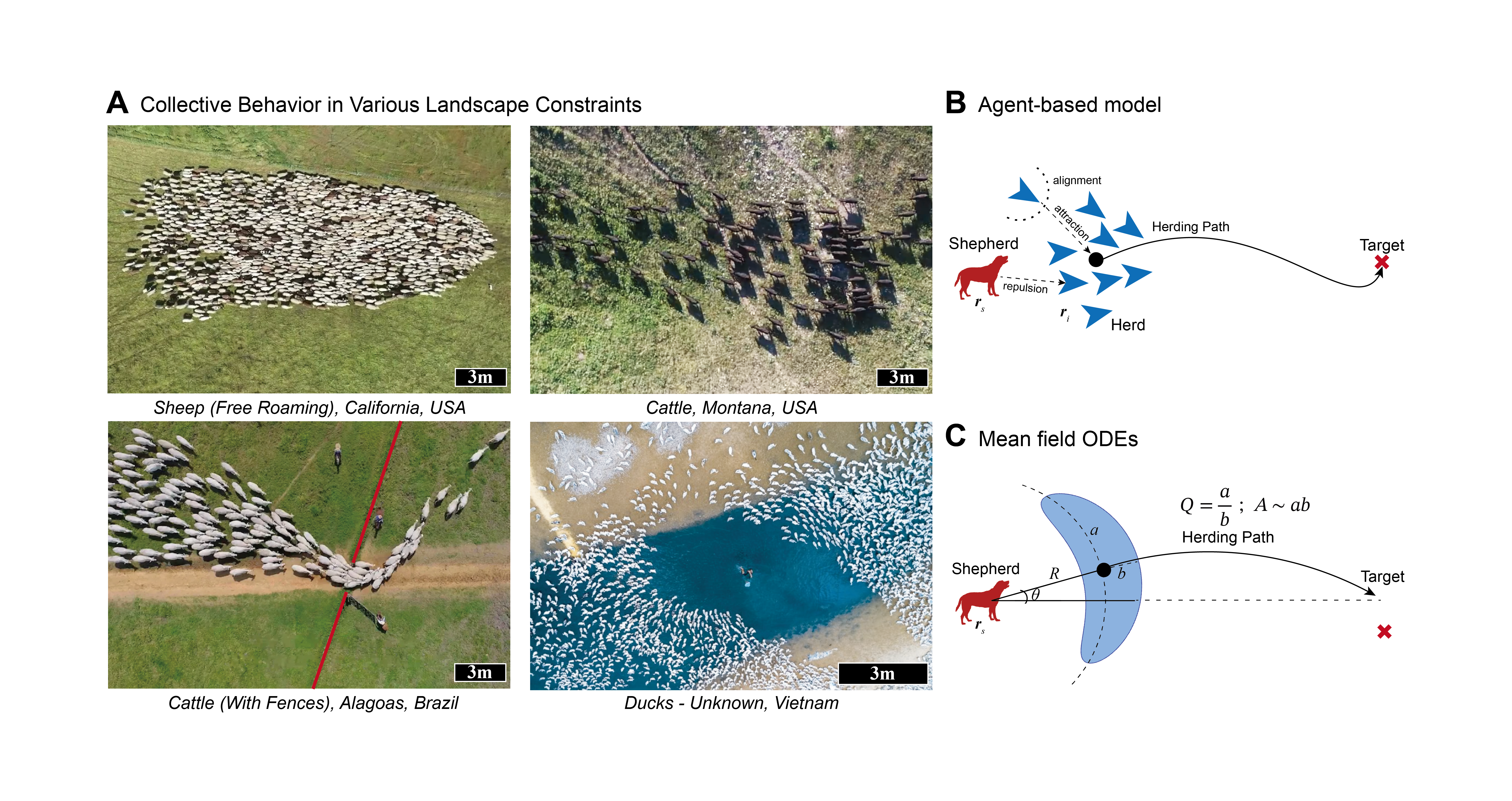}
\caption{Herding and models for it. (A) A herd of sheep, cattle, and ducks. (B) In our agent-based formulation, the position and orientation of individual herd members (blue arrows) respond to each other and the location of a nearby dog (red) that moves so as to optimally transport the herd center (black dot) to the target location (red cross). (C) In our mean-field ODE formulation, the herd is modeled as an ellipse with area $A$ and aspect ratio $Q$ that evolve in response to the distance $R$ to the dog. Cattle from Montana images courtesy of Jordan Kennedy. Ducks image by Cao Ky Nhan (https://mindthegraph.com/blog/photography-contest/). Free roaming sheep image from CBS (https://www.youtube.com/watch?v=0yMmc2xmE8I). Cattle with fence image from BBC (https://www.bbc.com/reel/video/p07t7zbv/turning-old-pastures-new).} 
\label{fig:1}
\end{figure*}

The visibly beautiful and complex self-organized patterns by which social organisms establish and maintain flocking behavior have led to a flurry of research activity over the last few decades \cite{reynolds1987,vicsek1995,sumpter2010, vicsek2012,bechinger2016active, berdahl2013emergent, gompper20202020, gomez2022intermittent}. The interest in understanding how flocks self-organize naturally leads to the question of how to control such a flock using a shepherd that can guide the herd in  biological, robotic, and other contexts \cite{strombom2014,bennett2012, falk2021learning, paranjape2018robotic, elamvazhuthi2020controllability, bongini2016sparse}. Here we explore the question of how emergent strategies arise from a simple optimization strategy that causes a shepherd to interact repulsively with a cohesive herd that it tried to move from one location to another. 
Complementing previous work that specifies explicit algorithms for herding \cite{strombom2014,bennett2012}, we show that distinct patterns of herding behavior resembling real-world behavior (Fig.~\ref{fig:1}A) arise naturally from a shepherd's attempts to simultaneously keep a herd coherent while moving it. A phase space characterized by the scaled size of the herd and the scaled speed of the shepherd shows the emergence of three herding strategies: driving, droving, and mustering consistent with qualitative observations in the field. Extensions of the model show that  the inclusion of inertial effects between agents in the herd does not change the qualitative nature of our results, and these strategies persist when a static target is replaced by a dynamic path that the shepherd is asked to guide the herd through. 

\section*{Mathematical framework}

\subsection*{Agent-based model}
We start with a population of $N$ self-propelled agents in two-dimensions, where they interact with each other following the Reynolds-Vicsek model \cite{reynolds1987,vicsek1995, couzin2002}. Accounting for the attractive and repulsive interactions within the herd, the finite speed of the flock, and the repulsive interaction between the shepherd and the agents being herded, the first-order dynamics of a given agent in the herd follows the equation of motion: 

\begin{equation}
\dot {\boldsymbol r_i} = v_a \left( \langle \hat{\boldsymbol v}_{i\neq j} \rangle _i -\gamma {\hat {\boldsymbol r}_{\rm cm-i}} +{\boldsymbol \eta} \right) -  \beta \frac{\partial U^a }{\partial {\boldsymbol r_i}} - \delta \frac{\partial U^s}{\partial {\boldsymbol r_i}} 
\label{herding::eqn:dynamics}
\end{equation}
An alternate second-order (inertial) model for the agents and the sheep discussed later in our study shows that the qualitative behaviours of the two models are similar. Here $i \in [1,N]$ and $N$ is the total number of agents in the flock, $v_a$ is the characteristic speed of agents in the flock sans shepherd, $\gamma$ sets the strength of the agent--agent attraction, and $\beta, \delta$ set the strengths of the repulsive agent--agent and agent--shepherd potentials, $\eta$ is a uniformly distributed additive noise corresponding to errors of perception or movement. The first term in the RHS of Eqn.~\ref{herding::eqn:dynamics} captures the local alignment between agents, where $\langle \hat{\boldsymbol v}_{i\neq j} \rangle _i$ is the average orientation of $i$-th agent's neighbors within an interaction radius $\Delta r < r_{\rm alignment}$. The $\gamma$ term in Eqn.~\ref{herding::eqn:dynamics} represents the attraction of an agent toward its (local) center of mass, ${\hat {\boldsymbol r}_{\rm cm-i}}$, where the averaging is taken over some cutoff radius $\Delta r < r_{\rm attract}$. In the large cutoff limit,  $r_{\rm attract} > \sqrt{N} l_a$, this term tends to a (weak) global attraction to the center of mass (CM) of the herd, which is the limit we consider (see SI for details). The gradient terms in Eqn.~\ref{herding::eqn:dynamics} represent radially symmetric repulsive potentials between any two agents, and between an agent and a shepherd, as follows: $U^a = E_0^a l_a \sum_i \sum_{j \neq i} \exp(-r_{ji}/l_a)$ and $U^s = E_0^s l_s \sum_i \exp(-r_{si}/l_s)$, where $l_a$ and $l_s$ capture the effective agent-size and shepherd interaction length scales, and $E_0^{(.)}=1$.

The shepherd is assumed to move at a constant speed $v_s$ and can control its location, ${\boldsymbol r}_s(t)$, by changing directions to transport all agents from one location to another. We assume that the shepherd continuously aims to reduce the herd's distance to the target, while also serving to keep the herd confined, while also reducing the shepherd's own deviation from the line connecting the target and the herd's (current) CM, suggesting the use of an optimizing objective function of the form

\begin{equation}
    C(\boldsymbol{r}_s; \boldsymbol{r}_{\rm target}, \boldsymbol{r}_a) = W_{\rm mean} |\boldsymbol{\Delta r_{\rm cm}}| + W_{\rm std} \sigma_{r_{cm}} + W_{\rm col} \Delta |R_{col}|, 
    \label{eqn:cost}
\end{equation}
where the first term, $|{\boldsymbol \Delta r}_{\rm cm}| = |{\boldsymbol r}_{\rm target}-{\boldsymbol r}_{\rm cm}|$ and ${\boldsymbol \Delta \boldsymbol r}_{\rm cm}$, represents the vector from the herd's center of mass to the target; the second term, $\sigma_{r_{\rm cm}} = \left [\frac{\sum_i (r_i- r_{\rm cm})^4}{N} \right]^{\frac 1 4}$, represents the importance of cohesion and is an estimate of the radial size of the herd; and the third term, $\Delta \boldsymbol{R}_{col} = {\boldsymbol r}_s - {\boldsymbol r}_{\rm cm} + l_s {\boldsymbol \Delta \hat{r}}_{\rm cm}$, represents the need to keep the flock within the line of sight of the shepherd (situational awareness) and penalizes the shepherd for being between the flock and the target (which would drive the flock away from the target). In other words, $\Delta \boldsymbol{R}_{\rm col}$ is minimized when the shepherd is directly behind the flock's center of mass along the line from the target to the herd's CM, such that ${\boldsymbol r_s}$, ${\boldsymbol r}_{\rm cm}$, and ${\boldsymbol r}_{\rm target}$ are all collinear. We used stochastic sampling at each timestep (see SI methods for detauls) to carry out a nonlinear optimization using the cost function above.

The results show the emergence of one of three distinct strategies for herding (Movie S1, S2). In the first strategy, the shepherd sweeps back-and-forth to continuously collect and compress the herd as it moves towards the target. This ``droving'' strategy has been documented in natural observations of herds \cite{sumpter2010} and also shown to be a feasible strategy for herding via simulations of shepherd interactions \cite{strombom2014}. In the first row of Fig.~\ref{fig:2}A, we see that it emerges naturally from our optimization procedure, and is not baked in. In the second strategy, the shepherd encircles the herd repeatedly as it moves towards the target. This ``mustering'' strategy  occurs when the shepherd and the herd move relatively quickly, as observed both in natural shepherding \cite{coppinger1993}. Again, we see that it emerges naturally for different relative sizes of the herd and the relative shepherd speed, as shown in the second row of  Fig.~\ref{fig:2}A. Finally, our simulations show the emergence of a third unexpected strategy, in which the shepherd is encircled by the herd and ``drives" it from within. In the bottom row of Fig.~\ref{fig:2}A, we show trajectories associated with this strategy that arise for still different relative sizes of the herd and the relative shepherd speed.

\subsection*{Low order model}

Before characterizing the regimes where we see these different strategies, we first consider a much simpler approach to the problem by coarse-graining the ABM and characterize the herd as a single entity that responds to a shepherd. Thus we aim to write down a low order dynamical system, e.g. a set of ordinary differential equations (ODEs) for the herd location using polar coordinates $R(t),\theta(t)$, the herd size parameterized by an area $A(t)$, and the shape of the herd $Q(t)$ (Fig. \ref{fig:1}C), characterized by its ellipticity.  In polar coordinates, given the radial dimension of the herd $r=R \pm b \sqrt{(1-(\phi-\theta)^2/(a/R)^2}$ (see SI for derivation) as shown in Fig.\ 1C, we see that the azimuthal width of the herd is $a(t)$ and the  radial thickness of the herd is $b(t)$. Then, at one extreme, the herd might have a lunate shape of angular width $\frac{a(t)}{R(t)}$, while at the other it is like a uniform ring of thickness $b(t)$, and in general, the area of the herd is given by $A(t) = \pi a(t)b(t)$  and its aspect ratio $Q = \frac{a(t)}{b(t)}$.

\begin{figure*}
\centering
\includegraphics[width=1.0\linewidth]{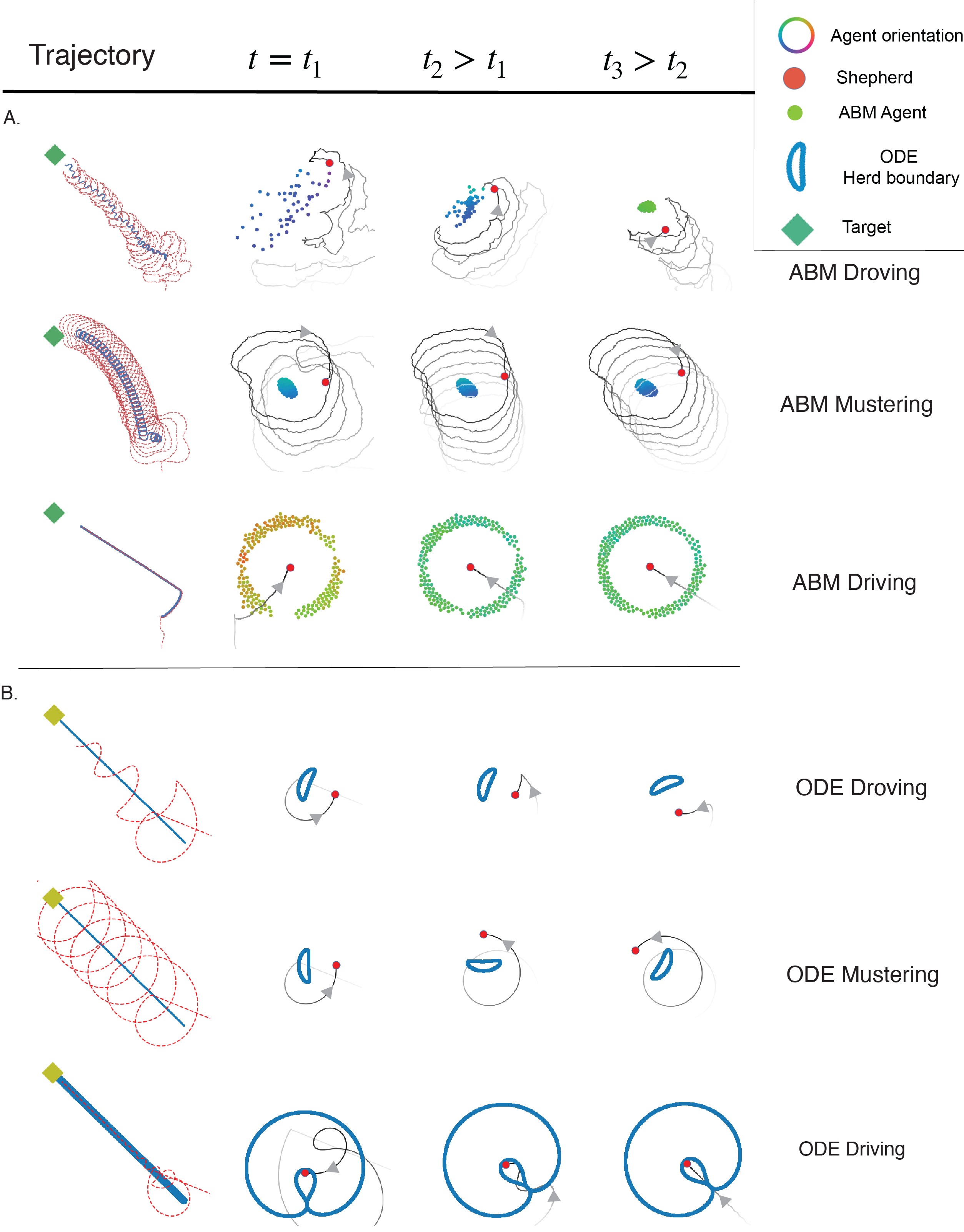}
\caption{Shepherd and herd trajectories across three regimes derived from the ABM (eq.(1-2)) and ODE (eq.(3-7)) models.   Trajectories correspond to movies S1 (A.) and S2 (B.) respectively (params in SI). In each of the six plots in the left column, the mean path of the flock (blue) over an interval is shown as it is driven by a shepherd on a separate path (red) towards a target (green square). Columns 2-4 show snapshots from column 1, with trajectories indicated in black, where fading indicates history. From left to right, snapshots represent the flock at later timesteps.}
\label{fig:2}
\end{figure*}

The dynamical equations for the location ($R(t), \theta(t)$), size ($A(t)$) and shape ($Q(t)$) of the herd in response to forces from the shepherd $f$ are given by

\begin{align}
\ddot R&=-\lambda_R (\dot R - u_R(t))+f, 
\label{eqn:Rddot} \\
\ddot\theta &=-\lambda_\theta[\dot\theta-Q^{-1}u_\theta(t)], 
\label{eqn:tddot}\\
\dot A&=-\lambda_A[A-A_0-\epsilon f + \zeta \frac{\bar {u_\theta} ^2}{R}], 
\label{eqn:Adot}\\
\dot Q&=-\lambda_Q[Q-Q_0-\omega f]. 
\label{eqn:Qdot}
\end{align}
Here the $\lambda_{(.)}$ set the inverse timescales of relaxation for the herd position, area, and aspect ratio, with $A_0$ and $Q_0$ being the relaxed areas and aspect ratios of the herd. Here $Q_0 = 1$ corresponds to a circular herd in the absence of a shepherd, $\epsilon$ and $\omega$ set the sensitivity of the flock's area and aspect ratio to the shepherd repulsive force which we take to be of the form $f=f_0e^{-R/l_s}$, where $l_s$ is the repulsion length-scale as before, and the shepherd's radial and angular velocity relative to the flock are given by $u_R$ and $u_\theta$, which are free to be manipulated by the shepherd. 
In the penultimate equation for the evolution of the herd size, we account for the observations seen in our numerical simulations that a shepherd encircling the flock tends to squeeze it periodically by using a phenomenological ``centrifugal" term of the form $\zeta \bar u_\theta^2/R$ which captures the radial squeezing effect.

Given the dynamical system (1.3-1.6), we then formulate an optimization problem similar to that for the ABM using a cost function 

\begin{equation}
    W_{\rm angle} |\theta(t) - \theta_{\rm goal}| + W_{\rm area} A(t) + W_{R} |R(t)-nl_s|
\label{eqn:ode_cost}
\end{equation}
reflecting the shepherd's aim to balance three goals: minimizing the angular deviation from the target direction $\theta_{\rm goal}$, minimizing the herd's size $A(t)$, and minimizing the distance $R(t)$ between the shepherd and herd's CM, where $n$ sets how far behind the herd's CM the shepherd wishes to be, with $n = 1$ corresponding directly to the collinear / line of sight term in the ABM model. We use a stochastic optimization procedure similar to that for the ABM model, with weights $W_{\rm angle}$, $W_{\rm area}$, and $W_{R}$ in eqn. \ref{eqn:ode_cost} chosen in analogy (but not in exact proportion) with those weights found in the ABM model,  and the condition $W_{\rm angle}  \gg  W_{\rm area} > W_{R}$ (see SI for parameters). In Fig.~~\ref{fig:2}B, we see that the ODE results match those obtained using the ABM model very well. 

\section*{Phase diagram of strategies}
\begin{figure*}
	\centering
	\includegraphics[width=1\linewidth]{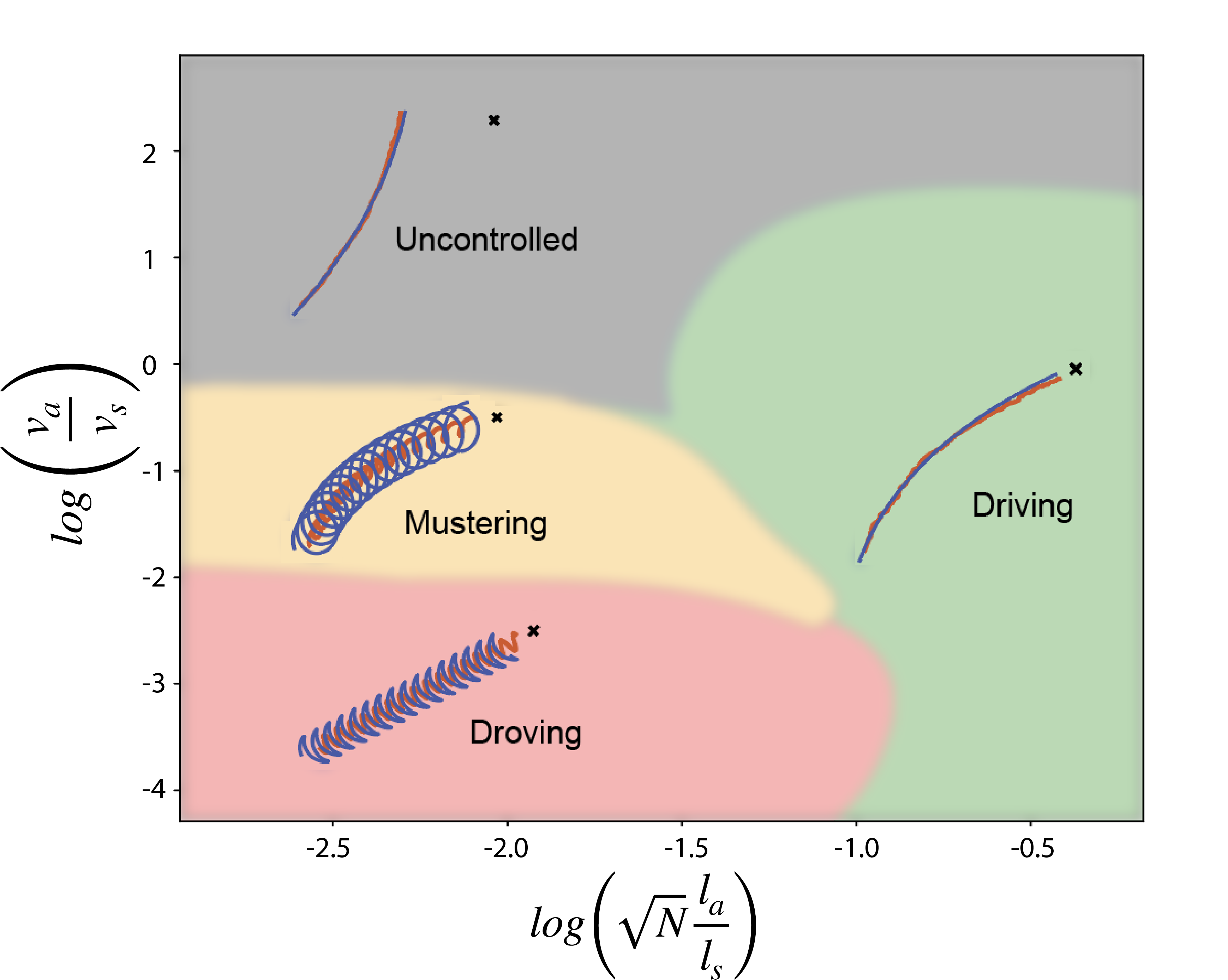}
	\caption{Phase plot of herding regimes, by solving the ABM model (eq. (1-2)). The shaded background corresponds to best fit predictions for regime boundaries using SVM (SI methods) trained on ~4000 ABM simulations, smoothed to reflect existence of overlap between regimes. For each combination of scaled swarm size (x-axis) and scaled swarm speed (y-axis), either the swarm could not be guided to the target with a single dog (gray), or it was most optimally guided with one of three herding strategy types: droving (red), mustering (orange), or driving (green). Overlayed on each regime is a sample agent-based trajectory of the center of mass of the herd (red line) and the center of mass of the shepherd (blue line), where the target location is represented by a black `x.' Parameters are specified in SI table 1.}
	\label{phase_diagram}
\end{figure*}

Having seen the emergence of the three strategies corresponding to droving, mustering and driving, we now explore their origin as a function of the phase-space of parameters corresponding to the scaled shepherd speed ($\frac{v_a}{v_s}$) and scaled herd size ($\frac{\sqrt{N} l_a}{l_s}$). In Fig. \ref{phase_diagram}, we see that when the herd is relatively small, droving or mustering are the preferred strategy: for faster shepherd velocities, droving emerges as the optimal, striking the best balance between maintaining herd cohesion while encouraging a high rate of travel of the herd. As the scaled shepherd speed decreases, there is a transition from droving to mustering, wherein a relatively slow shepherd must continuously strive to keep the herd together while incrementally moving the herd closer to the target. As the herd size gets larger, the relative speed at which mustering becomes dominant also decreases as bigger herds require even more effort to be encircled and kept together. For very large herds, mustering is no longer feasible and there is a transition to a driving strategy wherein the shepherd drives the herd from within it. In all cases, when the speed of the agents is much faster than the shepherd and/or the herd size is sufficiently large, the system becomes uncontrollable. 

\begin{figure*}
\includegraphics[width=1\linewidth]{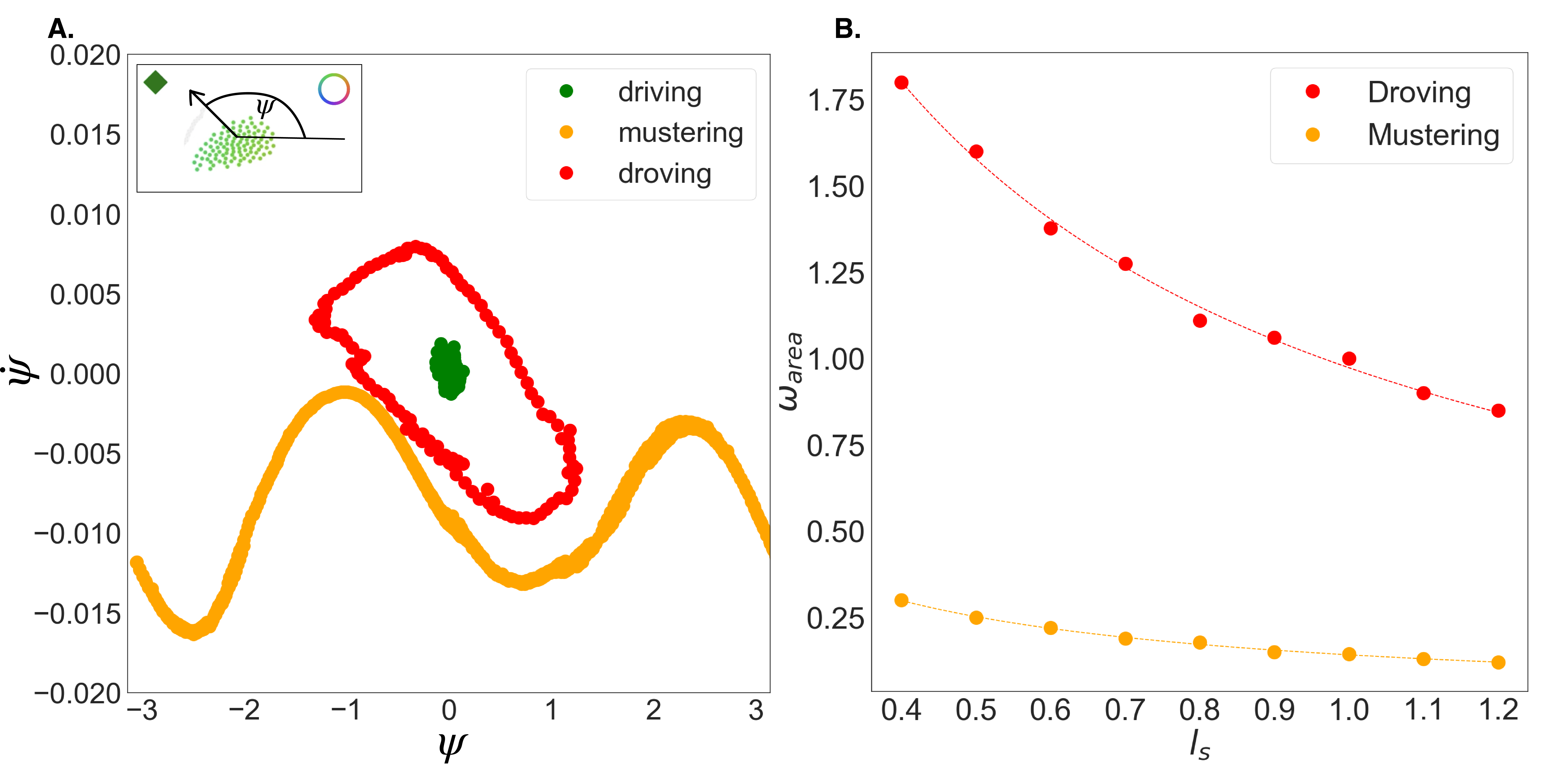}
\vspace{-.25in}
\caption{
 (A) Shepherding behavior (droving, driving, mustering) in herd orientation-angular velocity ($\psi$-$\dot \psi$) space, with data extracted from steady-state periods of the ABM (eq. (1-2)) simulations (Movie S1), and correspond to the different states of a simple pendulum (see text for details).(B) Plot of the shepherd-repulsion length scale ($l_s$) versus the herd's area compression rate ($\omega_{\rm area}$); dots  correspond to mustering and droving, respectively, and lines represent the theoretical fit to the scaling law $\omega_{\rm area} = c \frac{v_s}{\sqrt{N}l_a + d l_s}$, with $c=0.416, d=0.328$ for droving; and $c=0.046, d=0.706$ for mustering.}
\label{scaling_and_phasespace}
\end{figure*}

All successful strategies involve a combination of the herd/shepherd undergoing a simple combination of a translation and oscillation. In Fig.~\ref{scaling_and_phasespace}A we extract the oscillatory/rotational component of motion, plotting the herd's angular velocity $\dot \psi(t)$ as a function of the herd orientation $\psi(t)$, where $\psi = \langle \theta_i(t) \rangle$ represents the mean orientation of the agents in the herd. In the $\psi -\dot \psi$ plane, driving corresponds to a localized cloud centered at the origin, mustering leads to oscillations and droving leads to closed orbits. Not surprisingly, these modes of collective motion are analogous to the simplest dynamical system consistent with the inherent rotational symmetry, i.e. the different modes of a simple pendulum. Here the stationary fixed point corresponds to driving, oscillations correspond to mustering and librations correspond to droving. All these predicted behaviors are observed in our ABM and ODE simulations (Movie S1). 

Our simulations also show observations of oscillations of the herd's size during droving and mustering; these are due to the oscillatory motions of the shepherd around the flock, with a frequency $\omega_{\rm area}$. To derive an analytical estimate of $\omega_{\rm area}$ we note that the radius of oscillation of the shepherd scales as $\sqrt{N}l_a+d l_s$, where $d$ is a dimensionless constant,  $\sqrt{N}l_a$ is the size of the herd and  $l_s$ is the interaction length-scale of the shepherd. Thus, we expect $\omega_{\rm area} =c v_s/(\sqrt{N}l_a+d l_s)$, where $c$ is another dimensionless constant. Intuitively, this equation states that the period of oscillation is proportional to the distance the shepherd must traverse, or the circumference of the circle whose radius is a sum of the herd size and the shepherd repulsion length scale divided by the speed of the shepherd.  During droving, when the shepherd remains on one side of the herd, the shepherd spends less time oscillating around the herd, and we expect a higher frequency of oscillation $\omega_{\rm area}$. For mustering, we expect $\omega_{\rm area}$ to be lower. Both these predictions are consistent with the data as shown in Fig.~\ref{scaling_and_phasespace}B, and a good fit to our model for $\omega_{\rm area}$ is shown for both droving and mustering in Fig.~\ref{scaling_and_phasespace}B, indicating that the behavior of the herd can be predicted from the behavior of the shepherd, as expected.

All together, the variations in our results as a function of the ABM model parameters can be summarized as follows (with more detail and additional phase-diagrams in the SI): (I) reducing either the attraction strength $\gamma$ or the radius over which agents are attracted to their local CM $r_{\rm attract}$ leads to the diminished presence of driving in the phase diagram (Fig. \ref{phase_diagram}, SI Figs. \ref{SI_fig:attraction},\ref{SI_fig:attract_radius}), and when $\gamma \rightarrow 0$ or $r_{\rm attract} <l_a$, driving is no longer a possible solution. (II) Increasing the size of the agents in the herd, $l_a$, leads to enlarged mustering and driving regimes and a reduction in the prevalence of droving. This is expected as larger agents increase the relative size of the herd relative to the shepherd interaction length scale $\frac{\sqrt{N}l_a}{l_s}$ and also reduces the relative alignment radius $\frac{r^{\rm alignment}}{l_a}$. (III) Finally, decreasing the relative area-weight $W_{\rm std}$ in our objective function leads to the driving phase emerging at smaller herd sizes and the disappearance of droving and mustering, leaving a surprising uncontrolled region in the phase diagram for small herd sizes. Because herds subjected to driving are inherently less dense, it is reasonable to expect that a reduction in the emphasis on herd cohesion would make driving more favorable, though for small enough herds, driving is not an achievable phase.

\begin{figure*}
	\centering
	\includegraphics[width=0.9\linewidth]{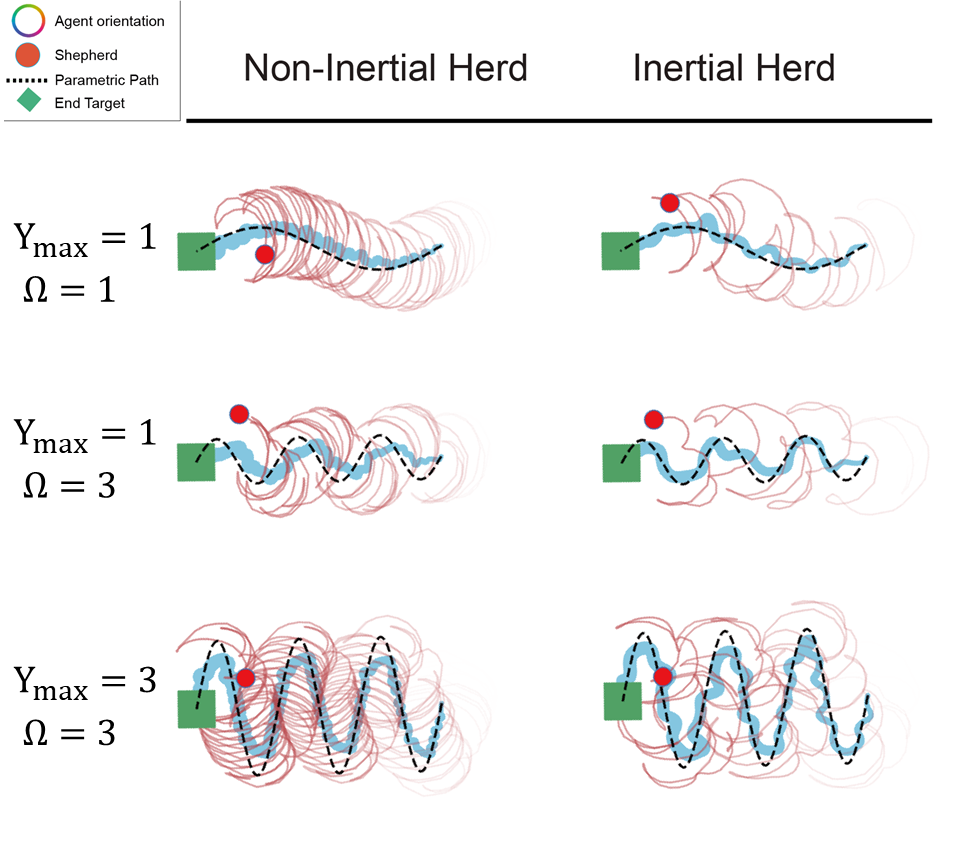}
	\caption{Comparing the strategies associated with non-inertial ABM simulations (eq. (1-2)) (column 1) and the inertial ABM simulations (eq. (\ref{fence::eqn:dynamics})) (column 2) when the target is moving according to a prescribed scheme ( eq.~\ref{herding::eqn:para_curve} -  $Y_{\text{max}}$ and $\Omega$ correspond to the amplitude and frequency  of the trajectory respectively). In each snapshot, the mean path of the flock (blue) over an interval is shown as it is driven by a shepherd on a separate path (red) over a parametric curve (black dashed line) towards the end target (green square).   We note only a minimal qualitative difference in herd trajectory when switching from an inertial to a non-inertial formulation.}
	\label{fig_5}
\end{figure*}

To further simplify and understand the emergent shepherd behavior via the solution of an optimization problem, we now characterize the time-averaged behavior of the shepherding strategy via a parametric representation. Moving into rotating frame (w.l.o.g.) so that the mean velocity of the CM of the herd $\bar v$ is along the $x$-axis, the mean shepherd motion may then be written as  $x_s(t) = R_x \cos(\omega t) + \bar v_s t$ and $y_s(t) = R_y \sin(\omega t)$, i.e., a superposed oscillatory component with a frequency $\omega$ and amplitudes $R_x$ and $R_y$ along, and normal to, the direction of motion. In the droving regime, $R_x \sim 0, R_y \neq 0$, so that one has an oscillatory velocity component perpendicular to the direction of motion. In the mustering regime, $R_x \approx R_y \neq 0$, and the shepherd continuously circles around the herd, leading to a low net drift speed of the herd toward the target. Finally, in the driving regime, $R_x \approx R_y \approx 0$. Observations of representative simulations (described in Movie S1) for different (scaled) herd sizes and (scaled) shepherd speed indicate that $\bar v^{\rm droving} \sim \bar v^{\rm driving} \gg \bar v^{\rm mustering}$. Rescaling this drift speed by the number of agents, we obtain an estimate of transport efficiency per agent, which is highest for driving, followed by droving and then mustering (see SI). 

Although driving exhibits higher efficiency in steady-state, it takes longer to reach steady-state. As expected, mustering trades off efficiency for herd compression. In all regimes, a significant amount of time is spent initially gathering and orienting the herd. Finally, it is worth noting that although strategies such as mustering require higher effort, they may offer auxiliary benefits such as the protection of a herd from an external threat or predator, and to favor this would require augmenting the cost function appropriately to account for factors that go beyond mere optimal transport of a herd. 



\section*{Model extensions}
So far, we have considered how an individual shepherd can optimally guide a non-inertial herd toward a target, in the spirit of generalizing the original Reynolds-Vicsek model towards functional tasks such as collective navigation. We now briefly discuss two extensions to the model that might be of value in a number of applications: (i) considering the role of inertial effects and (ii) considering guidance along a prescribed path.

\subsection*{Inertial effects}

Generalizing our first order ABM model into an inertial framework  leads to the equations of motion for the agents that now reads 





\begin{multline}
    \tau \dot {\bf v}_i = v_a \left( \langle \hat{\bf v}_{i\neq j} \rangle _i -\gamma {\hat {\boldsymbol r}_{\rm cm-i}} +{\bf \eta} \right) \\ -  \beta \frac{\partial U^a }{\partial {\bf r}_i}\hat{r} - \delta \frac{\partial U^s}{\partial {\bf r}_i} -\mu {\bf v}_i
    \label{fence::eqn:dynamics}
\end{multline}


where the terms are almost the the same as in Eqn.~\ref{herding::eqn:dynamics}. However,  Eqn. $\ref{fence::eqn:dynamics}$ differs from the first-order model via the addition of a response timescale $\tau$, and the addition of a dynamic viscosity term denoted by $\mu v$, where the coefficient $\mu$ that enforcing an acceleration penalty at higher speeds. When $\mu = 1$ and the agents are far away from each other, Eqn. \ref{fence::eqn:dynamics} simplifies to our original model treated in the previous part of the paper.

By simulating Eqn. \ref{fence::eqn:dynamics} using a RK4 numerical scheme and subject to the same optimization framework as for the first-order model, we observed qualitatively similar behavior (see Fig. \ref{fig_5}, Movie S3),  with all three herding strategies observed (albeit up to a shift in phase-space parameters), suggesting the robustness of the qualitative nature of the results using the simpler model.

\subsection*{Tracking moving targets}

A natural question that one can ask is if our approach can be generalized to allow for tracking/following dynamic targets with minimal modification. The inclusion of dynamic targets allows the shepherd not just to guide the herd to a target but along a path. In order to demonstrate the emergence of strategies in this more complicated scenario, we define a simple sinusoidal path along which the shepherd must guide the herd, which we define parametrically as follows:


\begin{equation}
\begin{cases}
    x_{\text{tar}}(i)&=i\cdot\frac{r_{\text{init}}}{i_{\text{max}}}\\
    y_{\text{tar}}(i)&=Y_{\rm max} \cdot\sin{(\frac{2\Omega \pi \cdot i}{i_{\text{max}}})}
\end{cases}
\label{herding::eqn:para_curve}
\end{equation}
 
where $i$ is a discrete time counter, with $i\in[0,i_{\text{max}})$ represents the $i$-th point in the discretized curved path. We set $i_{\rm max}$ to $500,000$ to obtain a smooth curve, and $r_{\text{init}}$ is the distance between the final target location and the herd's initial center of mass, i.e., $r_{\text{init}} = |\boldsymbol{r_{\rm herd}}(t=0) - \boldsymbol{r_{\rm target}}(i=i_{\rm max})|$, and $Y_{\rm max}$ \& $\Omega$ are the amplitude and frequency of the curve. In our simulations, the curve is stretched and rotated to fit the initial conditions and geometry of the system, such that $|\boldsymbol{r_{\rm herd}}(t = 0) - \boldsymbol{r_{\rm target}}(i=0)| = 0$. 

Within this formulation, we find that droving and mustering still emerge as optimal strategies for relatively small herds (see Fig. \ref{fig_5}) in the inertial and non inertial limit, though differences can be observed. However, not unsurprisingly, the shepherd trajectory is significantly reduced when agents in the herd are inertial, though the herd generally follows the prescribed path less accurately. We observe that inertial effects make the agents in the herd  move more cohesively with less short-time fluctuations since acceleration of the agents is penalized by inertia. However, this makes it harder for the shepherd to control higher order movements of the herd, but leads to less effort in controlling the general flow of the herd, as shown in Fig.~\ref{fig_5}. As a result, the inertial herd exhibits oscillations in motion (above the general sinusoidal motion of the path itself) that are not present in the non-inertial herd.

The consistency of strategies across static and moving target models suggests the optimization framework for shepherding outlined earlier is relatively robust.



\section*{Conclusions}
 
Inspired by observations of shepherding across the animal kingdom, here we have proposed and studied a minimal model of the dynamics of a shepherd guiding a herd towards a desired location or along a path. By investigating two complementary formulations of the underlying dynamics, we have shown the emergence of three fundamentally distinct families of optimal shepherding behavior---mustering, droving, and driving---that arise from our model in different parameter ranges, and characterized by a phase diagram of herding regimes (Fig.\ \ref{phase_diagram}) in terms of the scaled herd size $\sqrt{N} l_a/l_s$, and the scaled shepherd velocity $v_a/v_s$. Moving beyond the strategies to guide a herd to a target, we also considered the benefits and trade-offs in each strategy. A minimal model of efficiency measured by the time taken by the shepherd to transport a single agent show that droving is the most efficient way to transport small herds quickly while driving is most efficient for large cohesive herds. While our analysis is a step towards understanding herding strategies, it also raises a number of questions. Accounting for the effects of complex spatial geometry (e.g. the presence of boundaries and obstacles), shepherd training, and non-local strategies that sacrifice short-term control for long-term gain, the presence of multiple shepherds, and applications of shepherding to different fields \cite{turgut2008}  are all natural questions for future study.

\enlargethispage{20pt}

\section{Code Availability}
\vspace{-.2in}
Code to run agent-based and ODE simulations as well as create the phase diagrams listed in this paper is available on Github at github.com/arphysics/optimal-shepherding. Due to the large file sizes the outputs of all raw simulation (400-2000 simulations per phase diagram) are not listed on the Github but are available upon request.

\section{Author Contributions}
\vspace{-.2in}
A.R.: investigation, simulation, methodology, visualization, writing, review and editing. D.G.: simulation, visualization, writing. A.H.: methodology, visualization, review and editing. A.G.: investigation, methodology, writing, review and editing. L.M.: conceptualization, investigation, methodology, administration, resources, writing, review and editing. All authors give approval for publication.

\section{Acknowledgments}
\vspace{-.2in}
This work was supported by NSF GRFP (to A.R.), NSF Grants DGE-1144152 (to A.H.), DST-India/NSM/R{\&}D-HPC-Applications/2021/05 (to AG), NSF Physics of Living Systems Grant PHY1606895 (to L.M.), the Simons Foundation (to L.M.) and the Henri Seydoux Foundation (to L.M.) We are further grateful to members of the Mahadevan group past and present for the conversations that led to this work.



\newpage
\appendix

\renewcommand{\figurename}{Figure~A}
\renewcommand{\theequation}{A\arabic{equation}}
\setcounter{figure}{0}
\setcounter{equation}{0}

\section*{Appendix}

\subsection*{Movie description}
\subsubsection*{Movie S1--ABM MODEL}Example simulations of each herding regime in the agent-based model. \normalfont{Left: Full field view of the herded individuals (circles) and the shepherd (black cross, with grayscale trace) as it guides the herd to a target (green diamond). Right: View from the shepherd's frame. Marked as in left, but with continuous re-centering on the shepherd and re-rotating about the shepherd-target axis. (a) This video represents an example in which the shepherd is unable to control the herd. Notice that the small herd size reduces the cohesion between neighbors by limiting both local (Vicsek) alignment as well as long-range attraction. This means that a stable driving phase cannot be achieved and the system remains uncontrolled. (b) In this video, the shepherd primarily exhibits the side to side, `sweeping,' motion that is characteristic of the droving phase. On occasion, the shepherd fully encircles the herd, but the dominant behavior is to complete a half circle, ensuring that no agents escape the herd from the side, while still maintaining strong forward motion of the herd towards the target. In the droving phase, which is the most commonly observed phase in the real-world, a delicate balance between cohesion and forward motion has been achieved. This is only possible because the speed of the shepherd is high enough for it to react quickly to potential fractures in the herd. As the speed of the shepherd slows down, the system transitions to the mustering state. (c) In this video, the shepherd engages in a continuous mustering process to ensure that herd cohesion is maintained, while still generating some forward motion towards the target. In this scenario, the speed of the shepherd is not high enough to react quickly to small changes in herd behavior; the shepherd must consequently adopt a strategy that is more predictive than reactive. The mustering strategy offers such a process because it emphasizes a continuous effort to maintain herd cohesion as opposed to a more stop-start approach. (d) In this video, we see an example of transport in which the shepherd exerts control from within the herd, akin to a human controlling a car via the steering wheel. This type of behavior allows a shepherd even of moderate speed to control a very large herd around which it could not muster or drove. In this phase, the shepherd emphasizes small changes in response to the sheep in front of it versus the sheep behind it (corresponding to the line of sight term in the objective function). These small variations give a directionality to the motion allowing the shepherd to push (from within) a small group of agents towards the target. The collective motion is then ensured by the long-range attraction between agents and Vicsek alignment. Of further note: steady-state timeframes for each simulation in S1 are the following: Droving:400-600; Mustering: 6000-7000; Driving: 800-1000. These timeframes correspond to those shown in the main-text figure \ref{scaling_and_phasespace}.}

\subsubsection*{Movie S2--ODE MODEL} Example simulations of each herding regime in the ODE model. \normalfont{Full field view of the herd boundary (blue) as the herd is guided rightward by a shepherd (black circle). Left-hand-side panel shows the system in the `lab-frame.' Since the ODE simulations are solved in the frame of the shepherd (i.e., the shepherd's eye view), lab-frame videos are generated by first translating the shepherd-frame into the herd-frame and then translating the CM of the herd linearly to the target location (which is located in the same place as the ABM simulations, up to a scale-factor). The right-hand-side panel the behavior of the herd the original shepherd frame in which the ODE equations are solved. Descriptions of each phase are the same as in Movie S1. Please note, however, that in the ODE simulations, no information about the agents inside the herd is present. As a result, apparent rotations of the herd boundary only imply a change in the shape of the herd and do not imply that the agents within the herd are themselves `spinning.' }

\subsubsection*{Movie S3--Inertial ABM Model}
Example simulations of each herding regime in the agent-based model with inertia. Upper-Left: Full field view of the herded individuals (circles) and the shepherd (black cross, with grayscale trace) as it guides the herd to a target (green diamond). The cyan trace is the trajectory of the herd center of mass. Upper-Right: View from the shepherd's frame. Marked as in left, but with continuous re-centering on the shepherd and re-rotating about the shepherd-target axis. The bottom two panels are analysis related to individual agent's velocity. Movie S3 is a compilation of three scenarios that showcase we can recover the 3 phases (i.e. droving, mustering, and driving) when we introduce inertia to the model. (a) This is the droving phase characterized by the shepherd `sweeping' side to side while trying to push the herd to target, this is comparable to (b) in movie S1. (b) This part of the video showcases the `mustering' behavior of the shepherd. The main difference here is the shepherd's goal (parameter in cost function) is heavily weighted towards keeping the herd cohesive. This is comparible to (c) in movie S2. (c) We finally show, in this set of video, the `driving' case where the shepherd controls the group of agent within the herd. This is achieved through a balance of moderate shepherd speed, more center of mass attraction for the herd, as well as the shepherd focusing less on keeping the herd cohesive. This corresponds to (d) in movie S4. This is preliminary analysis for future work to showcase that these three strategies still emerge when inertia is introduced.

\subsubsection*{Movie S4--Paths \& Moving Targets (Non-inertial)}
Example simulations of each herding regime with the moving target scheme under non-inertial herd ABM model. Upper-Left: Full field view of the herded individuals (circles) and the shepherd (black cross, with grayscale trace) as it guides the herd along a path (end target shown in green diamond, the path shown in red, the current target following a bang-bang condition is showcased as a purple star). The cyan trace is the trajectory of the herd center of mass. Upper-Right: View from the shepherd's frame. Marked as in left, but with continuous re-centering on the shepherd and re-rotating about the shepherd-target axis. The bottom two panels are analysis related to current target moving speed and curvature. Movie S4 shows how a shepherd herd's the group of agent on a sine curve of changing amplitude/frequency. As the amplitude gets bigger, there are more straight line segments on the path and the shepherd makes less error whhile herding. 

\subsubsection*{Movie S5--Paths \& Moving Targets (Inertial)}
Example simulations of each herding regime with the moving target scheme under inertial herd ABM model. Upper-Left: Full field view of the herded individuals (circles) and the shepherd (black cross, with grayscale trace) as it guides the herd along a path (end target shown in green diamond, the path shown in red, the current target following a bang-bang condition is showcased as a purple star). The cyan trace is the trajectory of the herd center of mass. Upper-Right: View from the shepherd's frame. Marked as in left, but with continuous re-centering on the shepherd and re-rotating about the shepherd-target axis. The bottom two panels are analysis related to current target moving speed and curvature. Movie S5 shows how a shepherd herd's the group of agent on a sine curve of changing amplitude/frequency with inertia. In comparison to the non-inertial moview (S4), it seems that inertia can actually help the shepherd get the agents to target faster while sacrificing accuracy due to agents with inertia being harder to control.

\section*{Numerical Simulation and Optimization}


\subsection*{Initialization of simulations}
All agent-based simulations (ABMs) were initialized such that agents begin within a square box. The shepherd is always initialized along the y-axis at a distance greater than $l_s$ from the center of the herd. The target for all simulations was located such that the angle between the shepherd and the herd at initialization was not parallel, ensuring that a successful herding outcome required that the shepherd not just be able to propagate the herd to the target, but also cause the entire herd to change its orientation (which is a stricter test of the model.) The remainder of the parameters are described in table \ref{SI:TableABM}.

All ODE simulations were initialized with state values $R(0)=5$, $\theta(0)=0$, $A(0)=2$, and $Q(0)=1$. Only the parameters were chosen to differ between simulations, as listed in Table S3. Simulations were run with a time step of 0.05 and finite-difference updating.

\subsection*{Numerical integration of model equations}

To simulate our non-inertial ABM model formulation, we implemented a forward Euler timestepping algorithm in C++ to numerically integrate each system of model equations. The forward Euler method was chosen because the ABM model describes a first-order system in which the velocity of the shepherd is directly chosen by the shepherd. Additionally, a forward Euler approach most closely resembles the process of an agent making a decision, taking a step, and repeating the process. The behavior of the shepherd was determined by a discrete gradient descent approach at each timestep (corresponding to a `decision point' in the trajectory). At each step, several randomly selected directions, chosen from a uniform distribution between $[0, 2 \pi)$ were tested and the direction corresponding to the minimal objective function value was chosen by the shepherd. Classification of shepherd behavior into the different emergent strategies or phases was done upon completion of the simulation using a Python script.

Due to the simpler nature of the ODE model, the ODE simulations were implemented using a Runge-Kutta 4 scheme in Python, not C++, using the same discrete gradient descent approach for optimization described in the ABM model. For the ODE model formulation, we implemented a standard RK4 integration method due to the presence of some second order terms in the ODE equations. Given that the ODE simulation time does not scale as $N^2$ and is rather constant for any number of agents (because we only solve for the system area) the RK4 scheme's improved stability is well worth the speed and simplicity tradeoff in this formulation. Simulation plots were then created in two forms: the first form, shown in the right of all simulation videos (see Movie S2) represents the original ODE model from the shepherd-frame. The videos on the left show the entire ODE system placed onto a linear trajectory parallel to that observed in the ABM simulations to provide the reader a sense of what the lab-frame behavior of the herd-boundary looks like. While the lab-frame videos (Movie S2) provide an excellent depiction of the similarity between the ABM and ODE results, the exact trajectory and position of the target in the ODE lab-frame should not be interpreted literally.

Optimization was carried out by sampling $u_R, u_\theta$ in the range of $-u_R^{\rm max} \leq u_R(t) \leq u_R^{\rm max}$ and  $- u_{\theta}^{\rm max} \leq ,u_\theta(t) \leq u_{\theta}^{\rm max}$ (corresponding to moving towards the herd or around the herd) and chooses, at every timestep, the values of $u_R(t),u_\theta(t)$ that minimize the objective function (eqn. \ref{eqn:ode_cost}). The values of $W_{\rm angle}$, $W_{\rm area}$, and $W_{R}$ in eqn. \ref{eqn:ode_cost} were chosen in analogy (but not in exact proportion) with those weights found in the ABM model with $W_{\rm angle} \sout{\alpha} >>  W_{\rm area} > W_{R}$. 

\subsection{Phase space sampling}
The phase space of optimal shepherd behavior (Fig. \ref{phase_diagram}) was explored using a custom built bash script to launch $\sim 400-900$ simulations each time, automatically varying the speed of the shepherd $v_s$ and the number of agents $N$. This can be viewed in terms of a two-dimensional phase space corresponding to the scaled velocity $\frac{v_a}{v_s}$ of the agent (relative to the shepherd) and the scaled herd size $\frac{\sqrt{N} l_a}{l_s}$ (relative to that of an agent). Parameters for simulations were chosen to resemble real-world phenomena where possible. For example, based on evidence that herding dogs in the wild can control 10-100 sheep, we chose the number of agents in the simulations to initially be $N \sim 50$ \cite{king2012}. The speed ratio of the shepherd to the agents $\frac{v_s}{v_a}$ was controlled by setting the agents' grazing speed to a constant and varying $v_s$ which can be viewed as the shepherds `hunting' speed. Neither $v_a$ nor $v_s$ should be viewed as the true top speed of agents or shepherds in the wild. While simulations explored several orders of magnitude of the speed ratios, most simulations that show resemblance to real-world examples of hunting fall into a narrower window of speed ranges (see Fig. \ref{phase_diagram}).

The weights for the objective function were chosen on the assumption that losing agents in a herd was most unacceptable. Alignment was least important. As a result, the standard weights used in the objective function were: $ {W_{\rm mean}}/{W_{\rm std}} \sim 0.1$, and $ W_{\rm col} / W_{\rm std} \sim 0.001$.  

\subsection*{SVM method for classification}
Classification of region boundaries (mustering, driving, droving, uncontrolled) was done via a SciPy based \cite{scipy_SVM} support vector machine (SVM) with an RBF (radial basis function) kernel and $\gamma$ of $0.5$. This approach was chosen because an SVM based classification overlay provides a blend of simplicity in classification while maintaining a predictive ability. The SVM was trained on a dataset of approximately 2000 simulation runs of the agent-based model for each phase diagram.

\subsection*{Numerical optimization for model selection}

The trajectory of the shepherd was optimized on a timestep by timestep basis, via a discrete gradient-descent style sampling for both the ABM simulations and ODE simulations. Parameter values are listed in Tables S1-3. 




\section*{Simulation details, and Parameter Tables}

\subsection*{Typical herd sizes of various herding animals} An important parameter in our model is the number of agents $N$ that comprise a herd, and a key finding from this model is that the optimal shepherding strategy can vary based on this value (Fig.\ 3, Table S1). Typical herd sizes vary dramatically between different herding animals and geographic regions, and hence our model suggests that different shepherding strategies are useful in different herding contexts.

For example, dairy cattle herds typically fall in the wide range of 100 to 200 cattle per herd in the United States, a value which has risen over the past few decades; in contrast, dairy cattle in New Zealand and Australia form much larger herds of ~400 and ~250 cattle, respectively, while Germany typically sees a more modest 30-50 cattle per herd. \cite{cattle}. Sheep flocks in the United States are even more variable, sometimes consisting of fewer than 20 ewes in one flock, but also sometimes consisting of more than 500 ewes in one flock \cite{sheep}. Bird flocks, while less relevant to our work due to their three-dimensional motion, show a similar pattern; the many bird species of Monterey County, California, for example, form flocks that span just two birds to several hundred \cite{birds}.

\subsection*{Classifying Shepherd Behavior in Frequency/Phase Space}
In addition to classifying the droving, mustering, and driving phases as previously described--in terms of the sweeping or circling motion of the shepherd--we can also describe each phase qualitatively and quantitatively in $\psi$-$\dot \psi$ space as shown in Fig. \ref{scaling_and_phasespace}a. In order to generate Fig. \ref{scaling_and_phasespace}a, $\psi$ represents the average vector orientation of the herd, such that

\begin{equation}
    Z = e^{i\psi} = \sum_i \frac{e^{i \psi_i}}{N},
\end{equation} 

where $\psi_i$ represents the orientation of each agent in the herd and the complex exponential is used to calculate the average orientation in order to avoid discontinuities associated with polar coordinates. In this manner, the angular velocity of the herd, $\dot \psi$ can also be derived such that 

\begin{equation}
    \dot Z = \frac{d}{dt} e^{i\psi} = i e^{i\psi} \dot\psi.
\end{equation}
Solving for $\dot \psi$ gives 

\begin{equation}
\dot \psi = \frac{\dot Z}{i Z},
\end{equation} 

which must be a real quantity and can be computed numerically using Euler's formula, a finite-difference calculation of $\dot Z$ and a little algebra. See `polarization\_plotter\_general\_e-i-theta.ipynb'* under code files for the exact details. *Note that the code refers to $\psi$ as $\theta$; we use $\psi$ in the paper so as to avoid any confusion with $\theta$ as it is used in the ODE model.

This way of characterizing behavioral modes is commonly used to describe oscillator systems and allows for a nice visual description of the oscillator behavior, as described in the Fig. \ref{scaling_and_phasespace}a. With these axes, our system takes on three distinct shapes, each corresponding to a particular behavioral mode: during stochastic steady state droving, behavior can be described by a circular shape in $\psi$-$\dot \psi$ space. This is because, as the shepherd sweeps from a value of $\psi_{\rm max}$ to $-\psi_{\rm max}$ ,where $\psi$ is taken to be the global angle between ($-\pi$, $\pi$), the rate of sweeping increases and decreases, spanning a range of  $\dot \psi_{\rm max}$ to $-\dot \psi_{\rm max}$. This behavior depicts a circle. Of course, our system being stochastic shows similarity to this circular behavior in regions of steady-state behavior, though not exact correspondence. Similarly, mustering behavior in which $\dot \psi$ is typically constant, would trace out a shifted line in $\psi$-$\dot \psi$ space. Once again, similarity with rotation to account for the translational motion of the herd are seen in Fig. \ref{scaling_and_phasespace}a. Finally, driving corresponds to the shepherd maintaining a (near) constant orientation, which should lead to a flat line along the x-axis in $\psi$-$\dot \psi$ space. This behavior is also observed in our model.

\subsection*{Exploring Scaling Behavior}
In the previous section, we described how the behavior of the shepherd in the droving and mustering regimes is analogous to an oscillator system. As such, a question emerges: what are the frequencies of oscillation that are predicted by the model, and how do these frequencies scale across the phase space described in Fig. \ref{phase_diagram}? In order to derive a prediction of the oscillation frequency, we observe that steady-state behavior in droving and mustering emerge when the frequency of periodic motion exhibited by the herd is in sync with the periodic `sweeping' or `encircling' of the shepherd. This leads to two natural timescales: first, the relaxation timescale associated with the compression and expansion of the herd in response to an external agent. More precisely this timescales corresponds to the timescale over which the herd area fluctuates as it is being simultaneously compressed and pushed towards the target; the shepherd's dual emphasis on these two goals leads to a periodic expansion and contraction of the herd area over time during steady-state behavior (or as steady-state as our stochastic model allows). The second timescale is that of the shepherd's oscillation, corresponding to the time the shepherd takes to complete either a complete `sweep' or `encircling' of the herd. 

In order to compare these timescales, we first generate a theoretical prediction for the sweeping timescale of the shepherd and then derive the actual herd relaxation timescales using a Fourier Transform of our agent-based simulations. Since the speed at which the shepherd moves $v_s$ is fixed in our simulations, we can derive a scaling law for the frequency of shepherd oscillation by considering the two length-scales and one velocity scale in our model. The first length scale, the size of the herd, is given by $\sqrt{N}l_a$. The second length scale is the characteristic length scale associated with repulsion to the shepherd, $l_s$. Thus, the effective radius of the herd is some linear combination of the two characteristic lengths. Adding in the shepherd velocity gives the following scaling prediction:

\begin{equation}
    \omega = a \frac{v_s}{\sqrt{N}l_a + b l_s}.
    \label{scaling_law}
\end{equation}

where $\omega$ represents the frequency, $v_s$ is the speed of the shepherd, $\sqrt{N} l_a$ represents the `size' of the herd and $l_s$ represents the interaction length-scale between the shepherd and the herd. Intuitively, this equation states that the period of oscillation is proportional to the distance the shepherd must traverse whose radius is a sum of the herd size and the shepherd repulsion length scale divided by the speed of the shepherd. The frequency is simply the inverse of the period.

\begin{figure}[!tbhp]
	\centering
	\includegraphics[width=\linewidth]{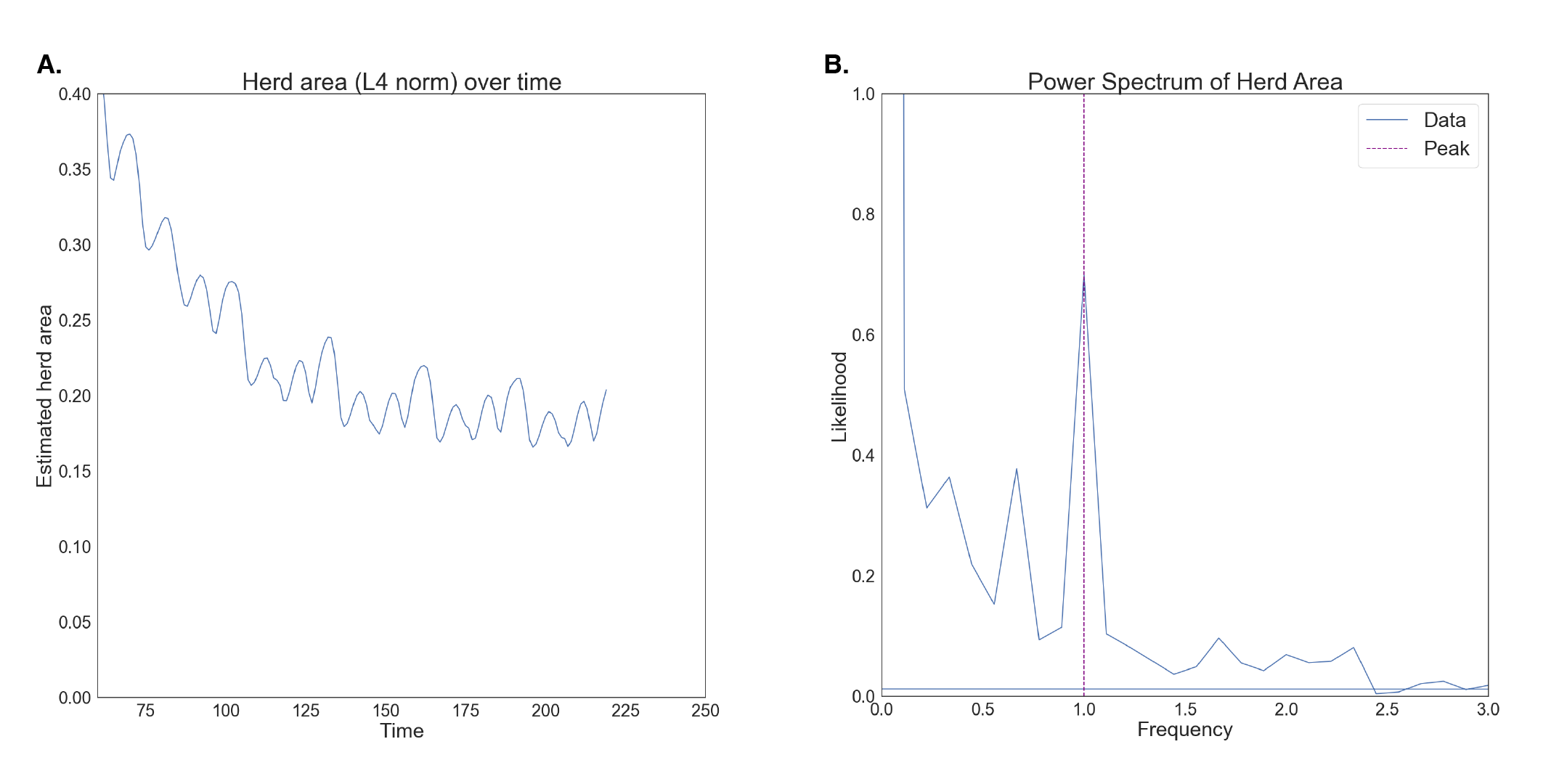}
	\caption{Fig. \ref{SI_fig:area_and_fft}A illustrates how the herd area fluctuates as a function of time during the mustering phase. Fig.  \ref{SI_fig:area_and_fft}B illustrates the fourier-transform of the area shown in Fig. \ref{SI_fig:area_and_fft}A. Solid blue line depicts the power spectrum of the herd;s area fluctuation; dashed vertical line shows the peak (corresponding to the primary oscillation frequency).}
	\label{SI_fig:area_and_fft}
\end{figure}

In order to extract the timescale and frequency of the herd area relaxation, we approximate the size of the herd using the $L_4$ norm as this is the same area approximating function used in the objective function during optimization (for the ABM model) (Fig. \ref{SI_fig:area_and_fft}A). We then convert the portion of this area time series into frequency space and look for the FFT peak (Fig. \ref{SI_fig:area_and_fft}B). Given that the data is stochastic, significant noise is present; while an automatic FFT power-spectrum peak detection algorithm could be used, in the interest of simplicity and given that only ~10 simulations were used to check the scaling law (which is sufficient because it spans a wide range of frequencies within each behavior phase), we chose to find the peaks in frequency space by hand.

Finally, in order to explore how our predicted and actual oscillation frequencies scale, we generated 9 simulations that span the left-right axis of Fig. \ref{phase_diagram} for both droving and mustering, by varying $l_s$. We chose to vary $l_s$ instead of $N$ or $l_a$ because large herd sizes relative to the shepherd repulsion length-scale leads the shepherd to adopt the driving behavior in which oscillatory behavior is negligible. As a result, we limited ourselves to the limit where $l_s > \sqrt{N} l_a$.


\subsection*{Herd boundary parameterization in our ODE framework}
We note that $A$, $R$, $Q$, $\theta$, $a$, $b$ are all dynamical variables and functions of time in our ODE model.  To capture a broad family of herd shapes ranging from elliptical to highly lunate, we define a herd shape in polar coordinates as described in the main text according to the equation

\begin{align}
\frac{(\phi - \theta)^2}{a^2}+\frac{(\frac rR-1)^2}{b^2} = \frac{1}{R^2},
\label{SI_eqn:ellipse}
\end{align}

If plotted in $r-\phi$ space, this shape would be always elliptical, but if plotted in Cartesian $xy$ space, this shape can take on a series of lunate forms. 

We note that an aspect ratio $Q$ can be computed as $\frac ab$, such that the major and minor axes can be defined as
\begin{align}
a=\sqrt{\frac{AQ}{\pi}},\quad
b=\sqrt{\frac{A}{Q \pi}},
\end{align}
With this definition, we have that
\begin{align}
\int_{-a}^{a}\int_{R-b\sqrt{1-(\theta/a)^2}}^{R+b\sqrt{1-(\theta/a)^2}}r\,\textup{d}r\,\textup{d}\theta=A,
\end{align}
as required for consistency.


\begin{figure}[!tbhp]
    \centering
    \includegraphics[width=0.5\linewidth]{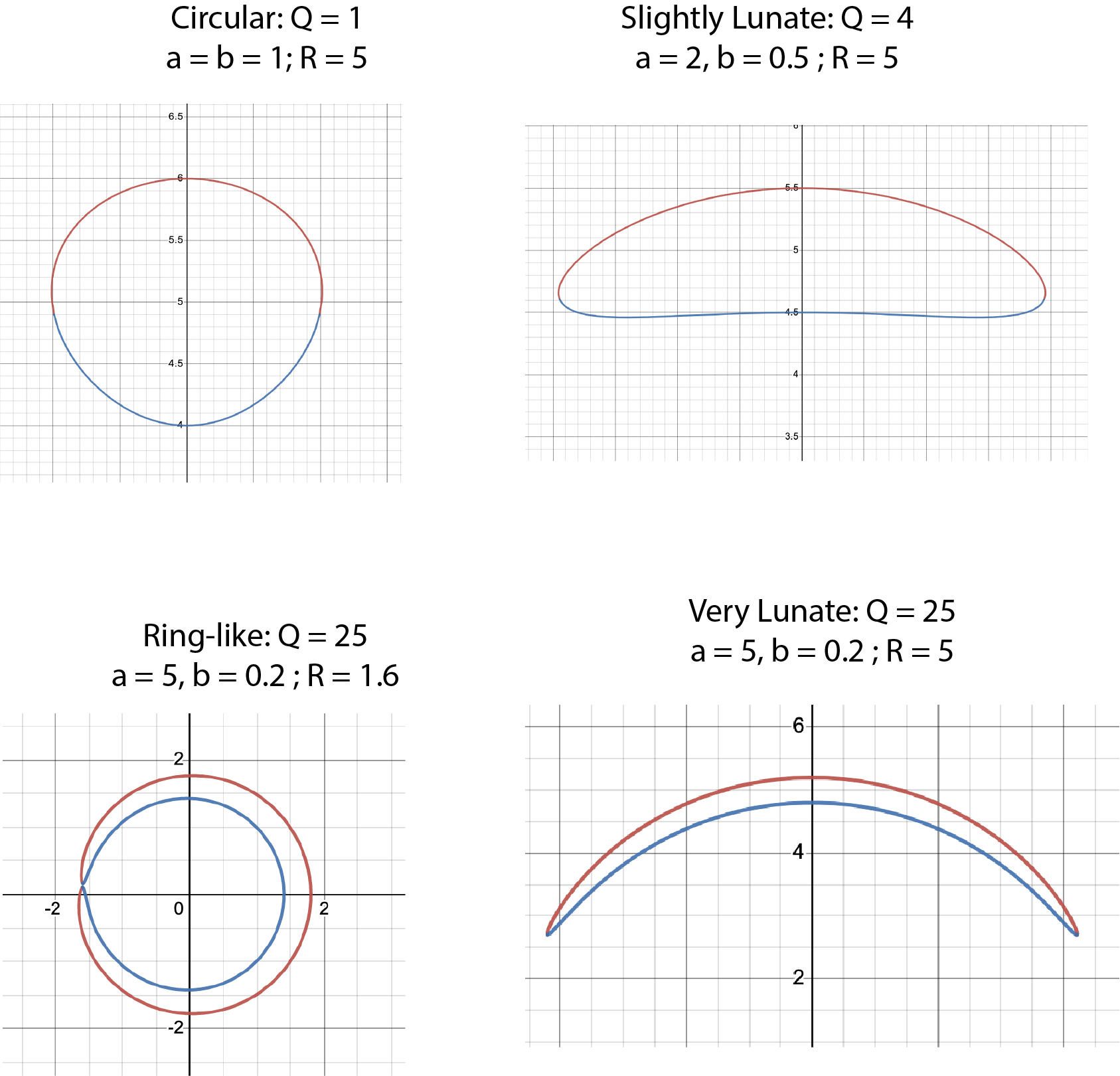}
    \caption{Illustration of the different herd shapes that can be achieved with our equation for the herd boundary, ranging from circular to lunate, to ring-like. The reader can play with the model if they so choose using the following link: \url{https://www.desmos.com/calculator/v7l33lt5k0}.}
    \label{SI_fig:HerdShapes}
\end{figure}




\subsection*{Parameter Dependencies}

\begin{figure}[!htb]
	\centering
	\includegraphics[width=\linewidth]{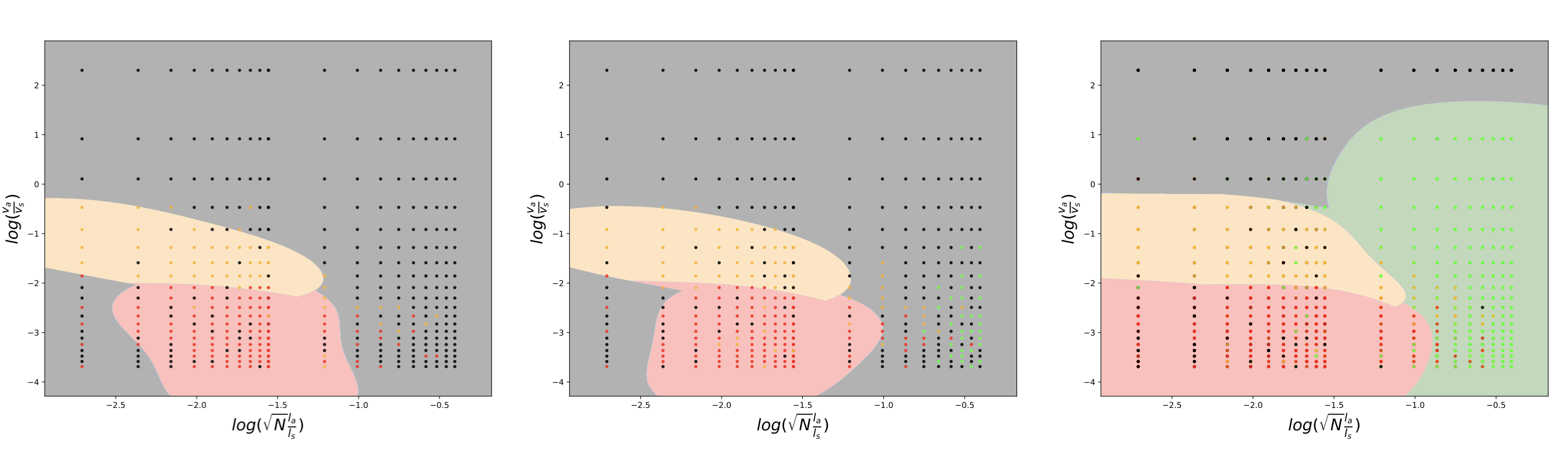}
	\caption{Phase plots of herding regimes, when we monotonically increase the weight of the center of mass attraction $\gamma$, (a) $\gamma = 0$, (b) $\gamma = 0.0005$, (c) $\gamma = 0.005$, which is a copy of figure~\ref{phase_diagram}. The rest of the parameters are the same as figure~\ref{phase_diagram}. We are using the same color code as figure ~\ref{phase_diagram}.}
	\label{SI_fig:attraction}
\end{figure}

\begin{figure}[!htb]
	\centering
	\includegraphics[width=\linewidth]{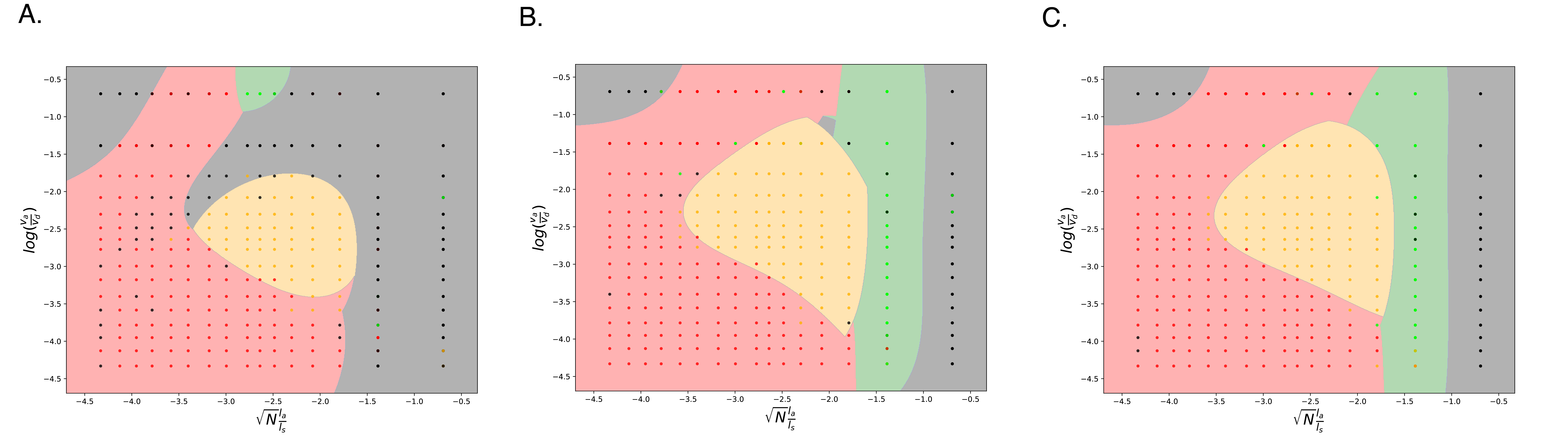}
	\caption{Phase plots of herding regimes, when we monotonically decrease the radius over which local attraction occurs $r_{\rm attract}$. (for N=100), the phase diagram resembles $r_{\rm attract}$. Note that in the limit of large $r_{\rm attract}$, (for N=100), the phase diagram resembles (b) \ref{SI_fig:N} and we recover the all three phases. As the radius of attraction decreases, the prevalence of the driving phase reduces, though driving can still be found, even when the attraction cutoff radius is on the same order as the Vicsek cutoff radius. Phase diagram labels correspond to the following: (a) $r_{\rm attract} = 10 l_a$, (b) $r_{\rm attract} = 50 l_a$, (c) $r_{\rm attract} = 100 l_a$.  We are using the same color code as figure~\ref{phase_diagram}.}
	\label{SI_fig:attract_radius}
\end{figure}



\begin{figure}[!htb]
	\centering
	\includegraphics[width=0.5\linewidth]{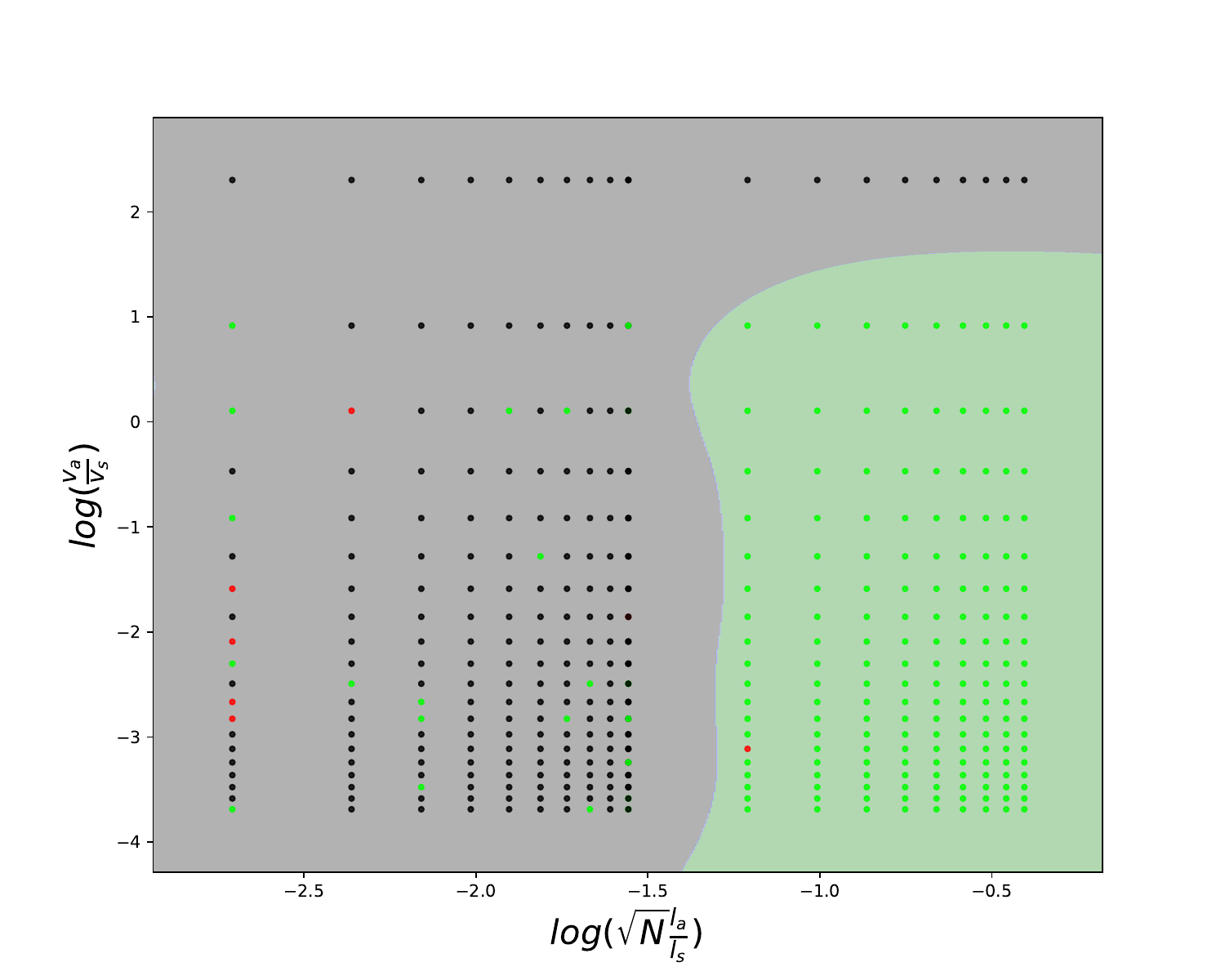}
	\caption{Phase plots of herding regimes, with a smaller weight of L4 norm (variance term) $W_{\rm STD} = 0.5$. In figure~\ref{phase_diagram} $W_{\rm STD} = 5$. The rest of the parameters are the same as figure~\ref{phase_diagram}, using the same color code.}
	\label{SI_fig:smallSTD}
\end{figure}

\begin{figure}[!htb]
	\centering
	\includegraphics[width=0.5\linewidth]{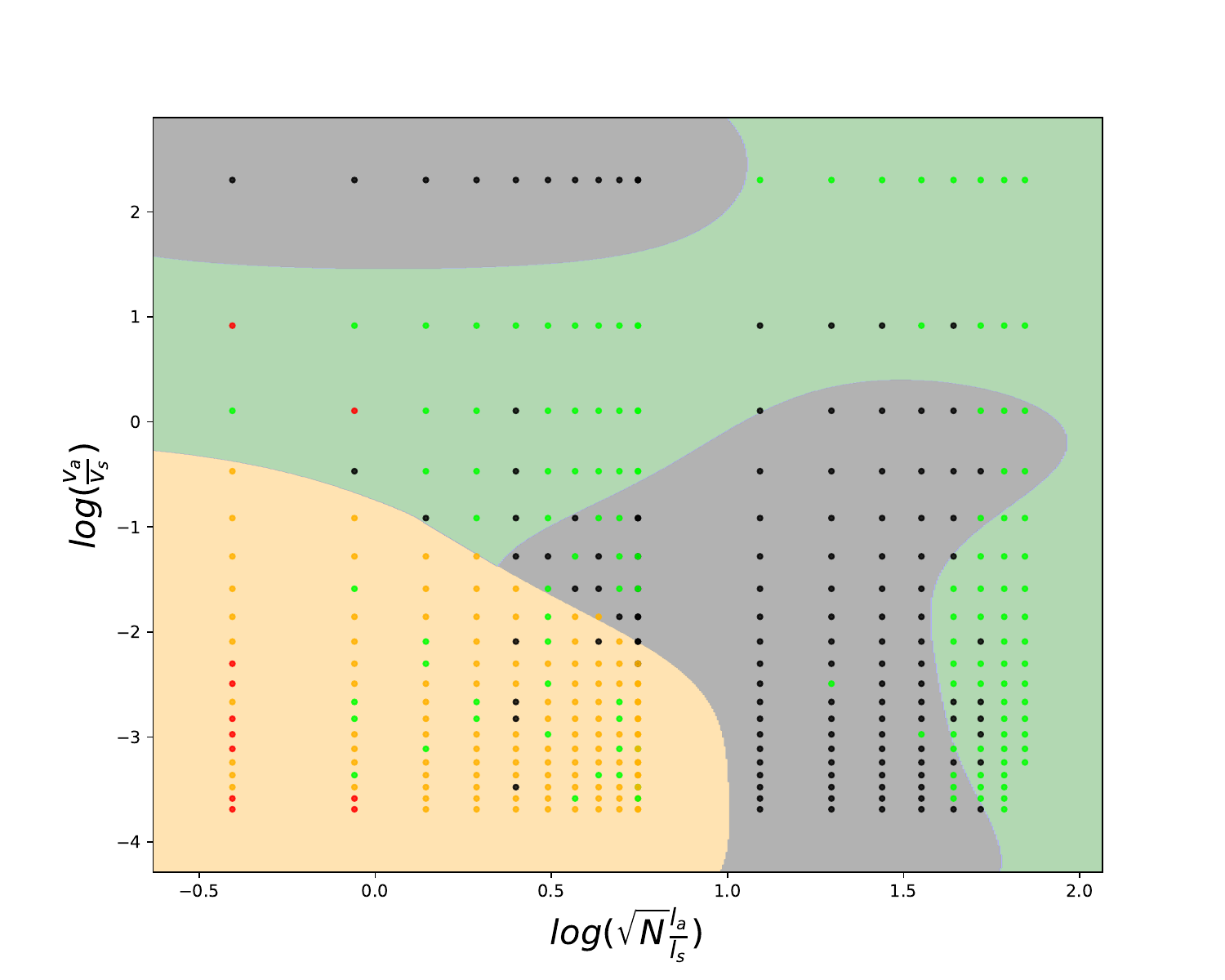}
	\caption{Phase plots of herding regimes, when we are using relatively bigger agents $l_a = 0.1$. In figure~\ref{phase_diagram} $l_a = 0.01$. Rest of the parameters are same as figure~\ref{phase_diagram}. We are using the same color code as figure~\ref{phase_diagram}. Color overlay: SVM $\gamma = 0.7$.}
	\label{SI_fig:largeAgents}
\end{figure}

\begin{figure*}[!htb]
	\centering
	\includegraphics[width=0.49\linewidth]{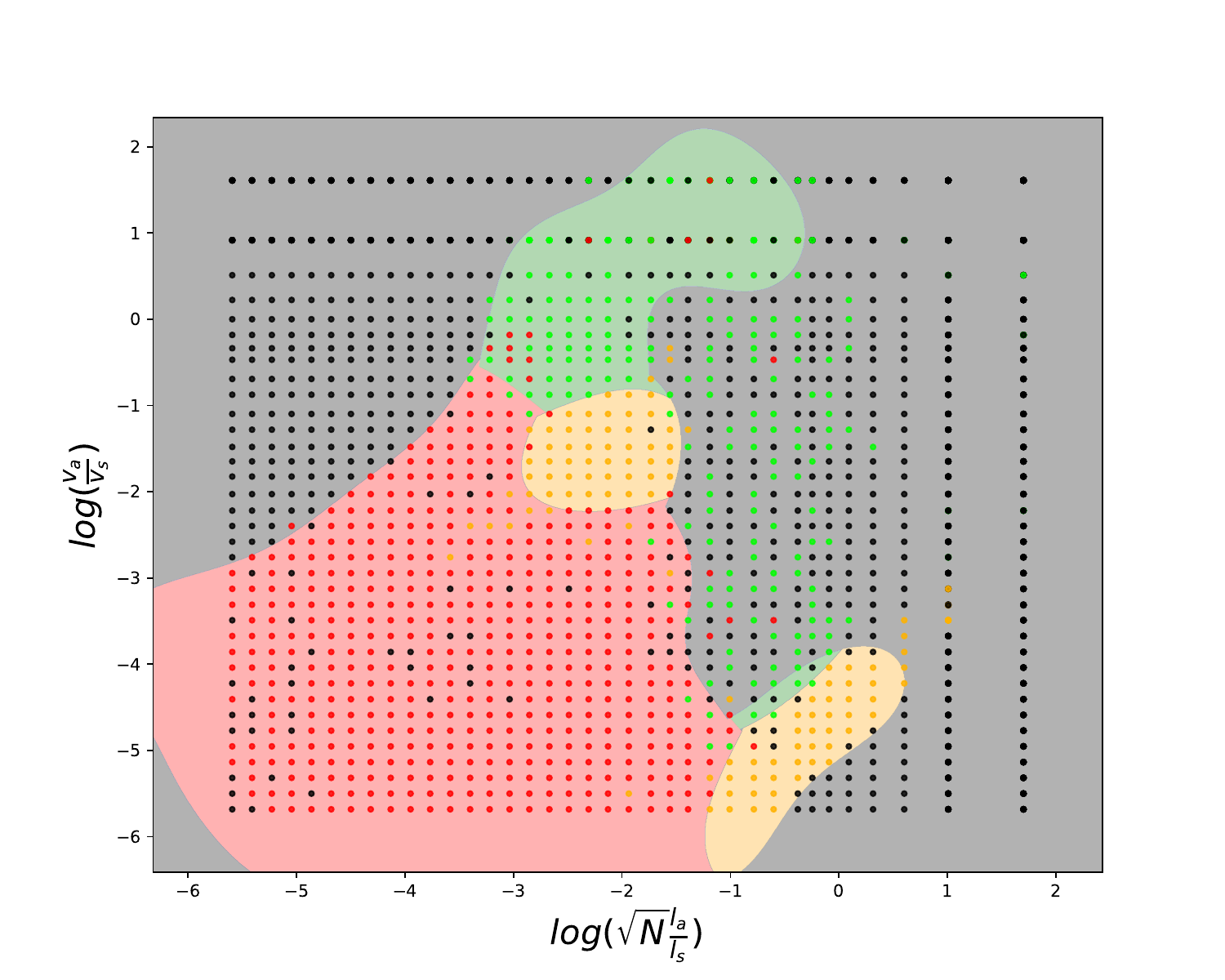}
	\includegraphics[width=0.49\linewidth]{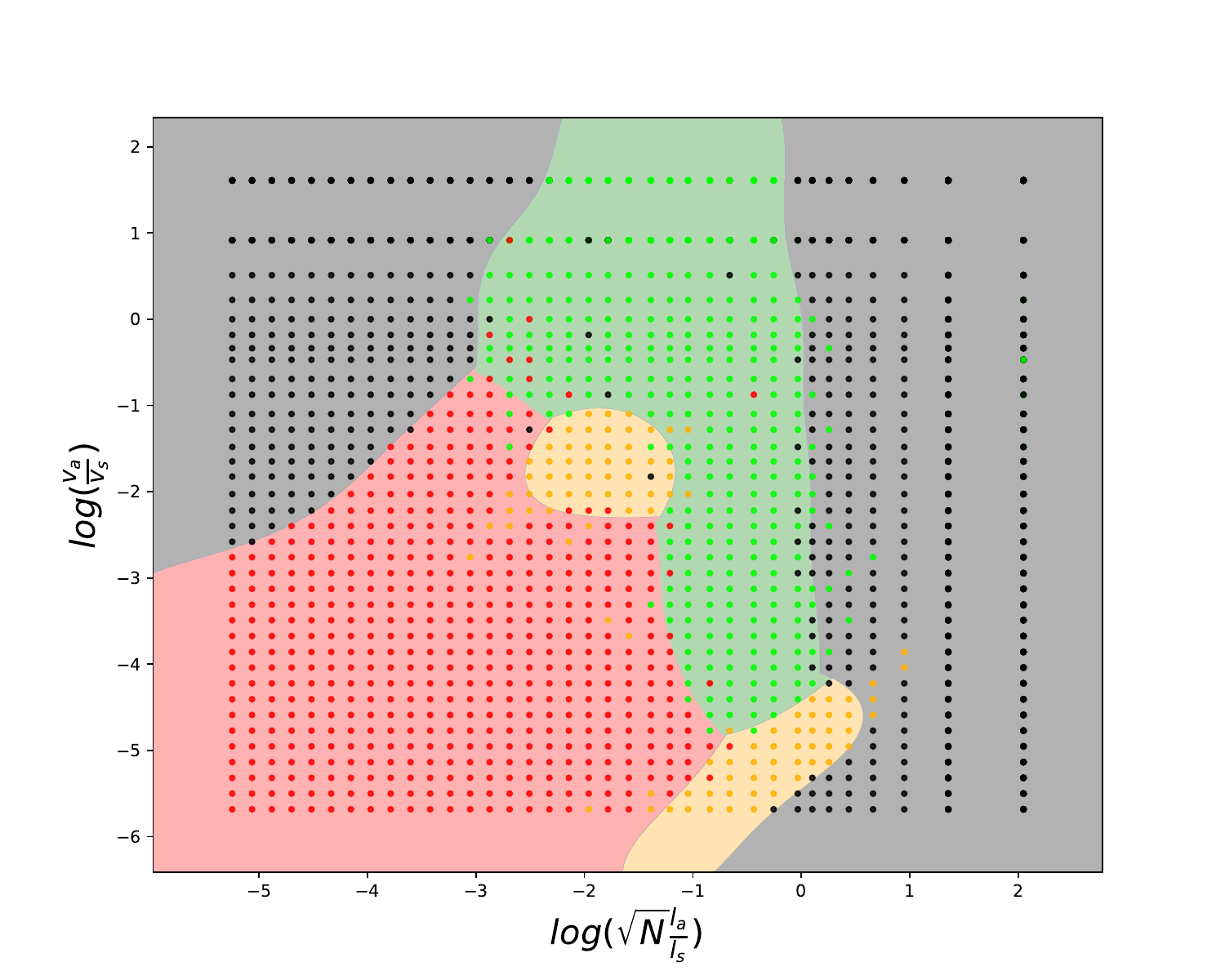}\\
	\includegraphics[width=0.49\linewidth]{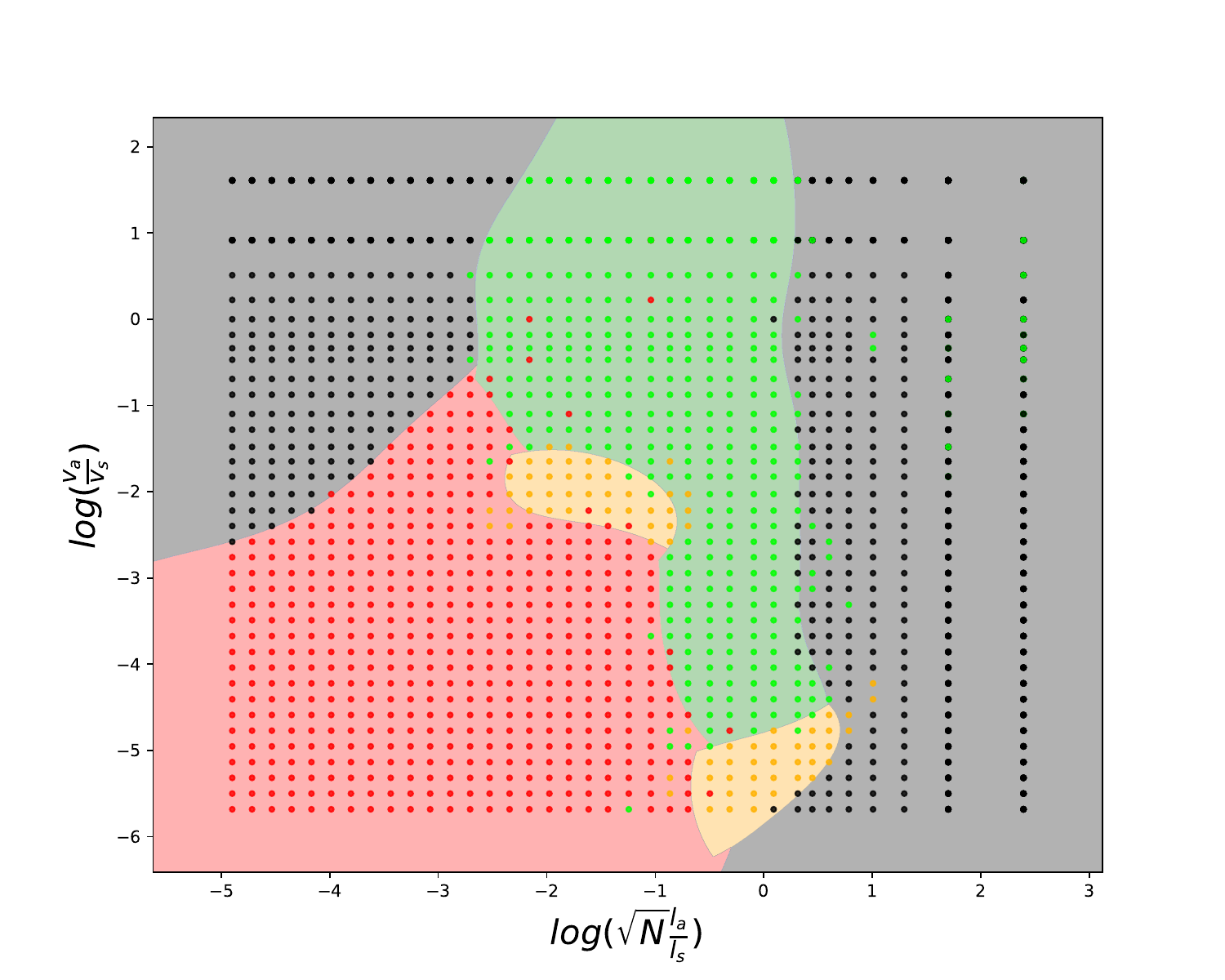}
	\includegraphics[width=0.49\linewidth]{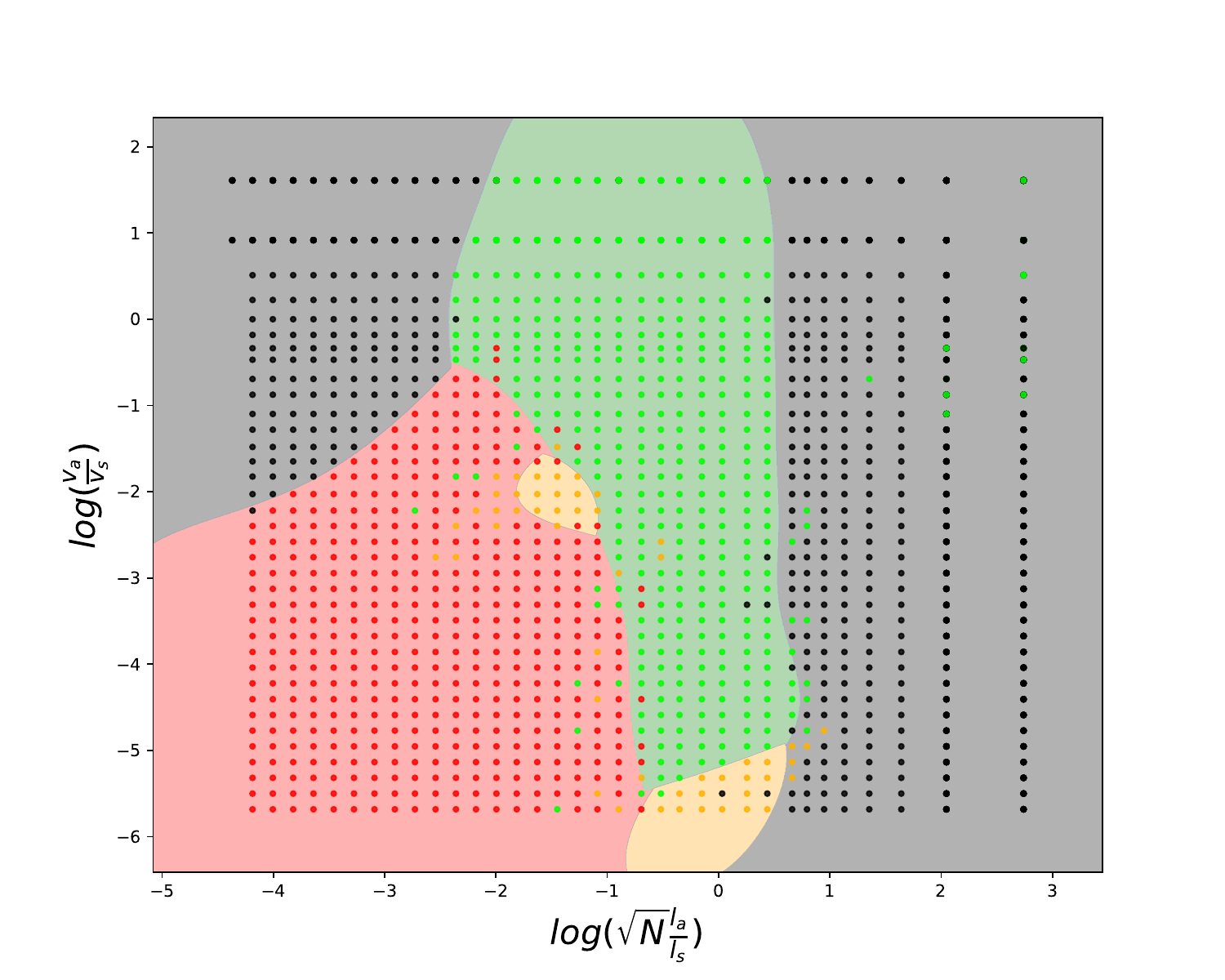}
	\caption{Phase plots of herding regimes while keeping the number of agents $N$ fixed (a) $N = 30$, (b) $N = 60$, (c) $N = 120$, and (d) $N = 240$ while varying the agent size $l_a$. Rest of the parameters are same as figure~\ref{phase_diagram}. We are using the same color code as figure~\ref{phase_diagram}.  }
	\label{SI_fig:N}
\end{figure*}

\begin{figure*}[!htb]
	\centering
	\includegraphics[width=0.5\linewidth]{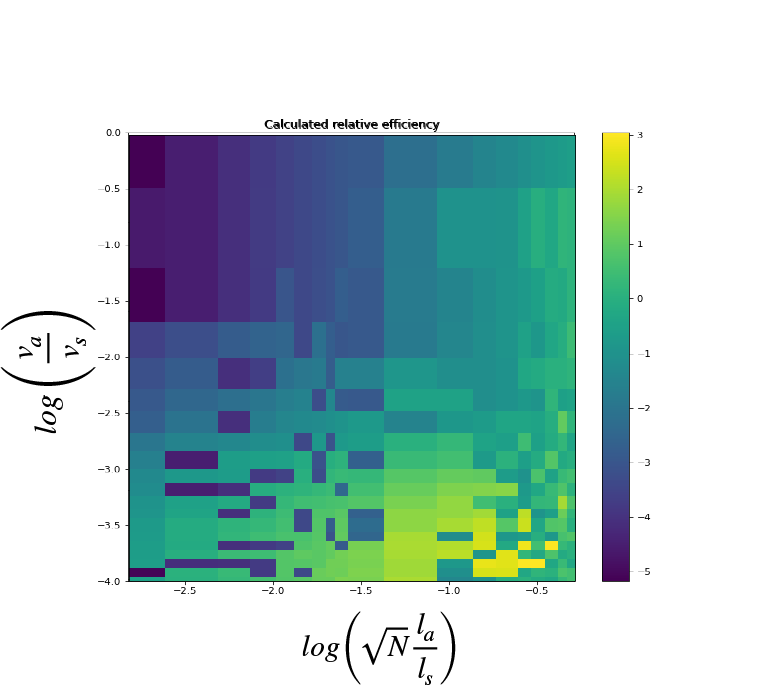}
	\caption{Phase plot of relative efficiency, defined as the fastest possible time to transport each agent to the target individually, divided by the actual transport time, i.e., $eff = \frac{N \Delta r_{cm}^{t=0}}{v_s \times T_{\rm actual}}$. High values indicate more efficient transport. Colorbar is on a logarithmic scale.}
	\label{SI_fig:efficiency}
\end{figure*}

\pagebreak

\begin{table*}[htb!]
	\centering
	\caption{Agent-based model parameters and typical values}
	\begin{tabular}{lllll}
	    \midrule
		Parameter Name        & Parameter Description                    & Droving & Mustering & Driving     \\
		\midrule
		$N$                   & Number of sheep                          &    50                  &   100                    &   200    \\
		$v_a$ 		          & Agent inherent speed                     &    0.05                &   0.05                   &   0.05  \\
		$v_s$                 & Shepherd speed                           &    1.3                 &   0.4                    &   0.2   \\
		$\ell_a$              & Agent size                               &    0.01                &   0.01                   &   0.01  \\
		$r^{\rm alignment} $  & Vicsek radius                            &    0.1                 &   0.1                    &   0.1   \\
		$l_s$                 & Shepherd repulsion length scale          &    0.3                 &   0.3                    &   0.3   \\
		$\alpha$              & Agent-agent alignment weight             &    0.1                 &   0.1                    &   0.1   \\
		$\beta$               & Agent-agent repulsion weight             &    0.1                 &   0.1                    &   0.1   \\
		$\gamma$              & Agent-agent attraction weight            &    0.005               &   0.005                  &   0.005 \\
		$\delta$              & Shepherd repulsion weight                &    0.9                 &   0.9                    &   0.9   \\
		$W_{\rm std}$         & Herd size weight (cost function)         &    5                   &   5                      &   5     \\
		$W_{\rm mean}$        & Target distance weight (cost function)   &    1                   &   1                      &   1     \\
		$W_{\rm col}$         & Line of sight weight (cost function)     &    0.001               &   0.001                  &   0.001 \\
		\bottomrule
	\end{tabular}
	\label{SI:TableABM}

	\centering
	\caption{ODE model parameters and typical values}
	\begin{tabular}{lllll}
        \midrule
        Parameter Name		& Parameter Description					& Droving	& Mustering	& Driving   	\\
        \midrule
        $l_s$			    & Shepherd repulsion length				&		1.5					&   1.5						&   1.5    					    \\
        $f_0$				& Shepherd repulsion weight				&		10					&   10						&   10							\\
        $\lambda_R$			& Translational drag frequency			&		0.5					&   0.5						&   0.5    						\\
        $\lambda_\theta$	& Angular drag frequency				&		0.05					&   0.05						&   0.05  						\\
        $\lambda_A$			& Area response frequency				&		10.0				&   10.0					&   10.0   						\\
        $\lambda_Q$			& Aspect ratio response frequency		&		1.5					&   1.5                   	&  1.5   						\\
        $A_0$				& Relaxed herd Area						&		5					&   5                    	&   180   						\\
        $\gamma$			& Flock size sensitivity to shepherd 	&		0.01				&   0.01					&   0.01  						\\
        $\omega$			& Flock shape sensitivity to shepherd 	&		5.0					&   5.0						&   5.0  						\\
        $\zeta$	& Flock area sensitivity to shepherd angular velocity &     0.01		        &   0.01					&   0.01 						\\
        $u_{\theta}^{\rm max}$& Max angular speed of shepherd       &       2.2                 &   2.4                     &   1.0                         \\
        $u_{R}^{\rm max}$   & Max radial speed of shepherd          &       0.5                 &   0.5                     &   0.5                         \\
        $W_{\rm angle}$   & Herd angle weight (cost function)          &       10.5                 &   10.5                     &   10.5                         \\
        $W_{\rm area}$   &  Herd area weight (cost function)         &       0.03                 &   0.03                     &   0.03                         \\
        $W_{\rm R} $   & Herd distance weight (cost function)          &       0.005                 &   0.005                     &   0.005                         \\
        \bottomrule
	\end{tabular}
	\label{SItable:ODEparams}
	
	\centering
	\caption{Non-dimensional parameters and values}
	\begin{tabular}{lllll}
        \midrule
        Nondimensional			&ABM 							&ABM 			&ODE 						&ODE			\\
        Param Description		&Param Definition				&Param Range	&Param Definition			&Param Range	\\
        \midrule
        Scaled herd size 		&$\sqrt{N} \frac{l_a}{l_s}$		& 0.002-20		&$\frac{\sqrt{A_0}}{l_s}$	& 0.33-120\\
        Scaled velocity 		&$\frac{v_a}{v_s}$				& 0.002-20		&--							& --			\\
        \bottomrule
	\end{tabular}
	\label{SI:tableNonDimParams}

	\centering
	\caption{Typical Parameter Values for Parametric Fit of ABM Simulations from Table \ref{SI:TableABM} }
	\begin{tabular}{lllll}
    \midrule
    Parameter Name		& Parameter Description					& Droving	& Mustering	& Driving    	\\
    \midrule
    $R_x$ & Oscillation length-scale || to $\bar v_s$ & 0.04 & 0.21 & 0.0 \\
    $R_y$ & Oscillation length-scale $\perp$ to $\bar v_s$ & 0.16 & 0.19 & 0.0 \\
    $\bar v_s$ & Average shepherd speed & 0.13 & 0.02 & 0.1 \\
    $\bar v_s N$ & Average agent transport rate & 6.5 & 2 & 20 \\
    $\omega$ & Shepherd Oscillation Rate & 4.78 & 0.5 & N/A \\
    \end{tabular}
    \label{SI:tableFitParams}

\end{table*}

\FloatBarrier

\pagebreak

\subsection*{Parametrized paths \& moving targets}



\subsubsection{Algorithm of Moving Target Implementation}

\begin{algorithmic}[!htb]
\If{$r_{\text{tar-max}}-r_{\text{cm}}>\kappa_{\text{upper}}\cdot l_s$} 
    \State $\text{flag}=0$ 
\Else{ $r_{\text{tar-max}}-r_{\text{cm}}<\kappa_{\text{lower}}\cdot l_s$}
    \State$\text{flag}=1$
\EndIf 
\If{$\text{flag}=0$}
    \State $\text{immediateTarget}_\text{x}=x_{\text{tar}}(i)$
    \State $\text{immediateTarget}_\text{y}=y_{\text{tar}}(i)$
\Else  
    \State $\text{immediateTarget}_\text{x}=\min{(x_{\text{tar}}(i+1),x_{\text{tar-max}})}$
    \State $\text{immediateTarget}_\text{y}=\min{(y_{\text{tar}}(i+1),y_{\text{tar-max}})}$ 
\EndIf
\end{algorithmic}



\section{Bibliography}
\bibliographystyle{unsrtnat}
\bibliography{refs}

\end{document}